\documentstyle[12pt,emlines,bezier]{article}
\newcommand {\bJ}{\mbox{\bf J}} 
\newcommand {\bQ}{\mbox{\bf Q}} 
\newcommand {\bj}{\mbox{\bf j}} 
\newcommand {\br}{\mbox{\bf r}}
\newcommand {\bd}{\mbox{\bf d}}
\newcommand {\bG}{\mbox{\bf G}}
\newcommand {\bv}{\mbox{\bf v}}
\newcommand {\bA}{\mbox{\bf A}}
\newcommand {\bB}{\mbox{\bf B}}
\newcommand {\bP}{\mbox{\bf P}}
\newcommand {\bM}{\mbox{\bf M}}

\newcommand {\bD}{\mbox{\bf D}}
\newcommand {\bH}{\mbox{\bf H}}
\newcommand {\bE}{\mbox{\bf E}}
\newcommand {\bq}{\mbox{\bf q}}

\newcommand {\bpi}{\mbox{\boldmath $\Pi$}}
\newcommand {\bppi}{\mbox{\boldmath $\pi$}}

\newcommand {\bT}{\mbox{\bf T}}
\newcommand {\bal}{\mbox{\boldmath $\alpha$}}
\newcommand {\bbe}{\mbox{\boldmath $\beta$}}
\newcommand {\bde}{\mbox{\boldmath $\delta$}}

\newcommand {\bp}{\mbox{\bf p}}
\newcommand {\ba}{\mbox{\bf a}}
\newcommand {\bk}{\mbox{\bf k}}
\newcommand {\bn}{\mbox{\bf n}}
\newcommand {\bR}{\mbox{\bf R}}
\newcommand {\cD}{\mbox{\boldmath $\cal D$}}
\newcommand {\cE}{\mbox{\boldmath $\cal E$}}
\begin{document}
\thispagestyle{empty}
\begin{flushleft}{\Large \bf MULTIPOLAR REPRESENTATION OF MAXWELL AND 
SCHR$\ddot O$DINGER EQUATIONS: LAGRANGIAN AND HAMILTONIAN FORMALISMS: 
EXAMPLES} 
\end{flushleft} \hskip 2 cm \vbox{\hbox{\bf 
V.\ M.\ DUBOVIK and B.\ SAHA} \hbox{\it Laboratory of Theoretical Physics} 
\hbox{\it Joint Institute for Nuclear Research, Dubna} \hbox{\it 141980 
Dubna, Moscow region, Russia} \hbox{\it e-mail:  
dubovik@thsun1.jinr.dubna.su} \hbox{\it e-mail:  
saha@thsun1.jinr.dubna.su} \vskip 3mm \hbox{\bf M.\ A.\ MARTSENYUK} 
\hbox{\it Department of Theoretical Physics} \hbox{\it Perm' State 
University} \hbox{\it 15 Bukirev Str., 614600 Perm' Russia} \hbox{\it 
e-mail: mrcn@pcht.perm.su}} \vskip 1.5cm 

\noindent 
Development of quantum engineering put forward new theoretical problems.
Behavior of a single mesoscopic cell (device) we may usually describe by
equations of quantum mechanics. However if experimentators gather hundreds
of thousands of similar cells there arises some artificial medium that
one already needs to describe by means of new electromagnetic equations.
The same problem arises when we try to describe  e.g. a sublattice 
structure of such complex substances like perovskites.
It is demonstrated that the inherent primacy of
vector potential in quantum systems leads to a generalization of the
equations of electromagnetism by introducing in them toroid
polarizations. To derive the equations of motion the Lagrangian and the 
Hamiltonian formalisms are used. Some examples where electromagnetic 
properties of molecules are described by the toroid moment are pointed 
out.

\newpage \section{Introduction} \setcounter{equation}{0} 
\vskip 3mm \noindent Let us remind a known thing that says {\it "it 
is impossible to introduce electrodynamics of matter in general"} ({\bf 
E.\ A.\ Turov, 1983}). For example, different types of crystalline 
structures of matter lead to the alignment of one or other type of 
polarizations in the matter considered. So the necessity to introduce in 
the equations of high tensor polarization is arose. But the most intricate 
case is the toroid polarization one. The fact is that the ideal static 
toroid moments do not interact with each other. So the dynamic 
alignment of toroid moments is impossible thanks to electric and magnetic 
interactions. But this takes place for example in perovskites [1,\,2] and 
have to be explained.  The other case when the local toroid moments can 
align is connected with their proper oscillations that permit them 
interact with each other [3] through electric and magnetic fields. 

It is noteworthy to remark that toroid moments are the multipole sources of 
field-free vector potentials. Therefore, electromagnetotoroidic equations 
we are forced to express in terms of vector potentials. The first part of 
our report is devoted to introduction of the equations mentioned in 
Lagrangian formalism. In the second part we introduce toroid moments in 
Schr$\ddot o$dinger equations.

\vskip .5cm

\section{Toroid moments in Lagrangian formalism} 
\setcounter{equation}{0}
\vskip 3mm

\subsection{Canonical formulation of electrodynamics}
\noindent

We begin with usual canonical formulation of electrodynamics (e.g. [4]).
The interacting system of electromagnetic field and non-relativistic 
charged particles is specified partly by a discrete set of variables, 
namely the coordinates of the charged particles, and partly by a 
continuous set, which we take to be the values of the vector potential in 
the Coulomb gauge. The Lagrangian $L$ will thus be a functional of ${\bA}$ 
and $\dot{\bA}$ if the particle coordinates and their velocities are 
fixed, and a function of the ${\bq}_\alpha$ and $\dot{\bq}_\alpha$ if the 
vector potential and its time derivative are fixed. We write 
$L\,=\,L[{\bA}, \dot{\bA};\,q,\,\dot q]$. In the application of Hamilton's 
principle, the particle and the field coordinates are to be varied 
independently. The Lagrangian must then be chosen so that variation with 
respect to the particle coordinates gives Newton's law $$ 
m_\alpha\,\ddot{\bq}_\alpha\,=\,e_\alpha\,\bigl({\bE}\,+\,\frac{1}{c} 
\dot{\bq}_\alpha \times {\bB}\bigr)$$
and variation with respect to the field coordinates (subject to the 
Coulomb gauge condition) gives the equation of motion
$$\nabla^2\,{\bA}\,-\,\frac{1}{c^2}\,\ddot{\bA}\,=\,-\,\frac{4\pi}{c}\, 
{\bJ}^{\bot}$$
for the vector potential.

A suitable Lagrangian is obtained by setting (see e.g. [4])
\begin{equation} L\,=\, L_{par}\,+\,L_{rad}\,+\,L_{int} \end{equation} 
with \begin{equation} L_{par}\,=\,\frac{1}{2}\,\sum_{\alpha}\, 
m_{\alpha}\,\dot{\bq}_{\alpha}^{2} \,-\,\frac{1}{2}\,\sum_{\alpha \ne 
\beta}\frac{e_{\alpha}\,e_{\beta}} {|{\bq}_{\alpha}\,-\,{\bq}_{\beta}|} 
\end{equation}
\begin{equation}
L_{rad}\,=\,
\frac{1}{8\pi}\, \int\, \bigl[\frac{\dot{\bA}      
^2}{c^2}\,- \,(\mbox{curl}\,{\bA})^2\bigr]\, dV
\end{equation}
and
\begin{equation}
L_{int}\,=\,\frac{1}{c}\,\int\,{\bJ}({\br}) \cdot {\bA}({\br})\, dV
\,=\,\sum_{\alpha}\frac{e_{\alpha}}{c}\,\dot{\bq}_{\alpha} \cdot 
{\bA}(q_\alpha, t). 
\end{equation}
Here $L_{par}$ is the Lagrangian appropriate to a system of charged 
particles interacting solely through instantaneous Coulomb force; it has 
the simple form of "kinetic energy minus potential energy". $L_{rad}$ is 
the Lagrangian for a radiation field far removed from the charges and 
currents, and has the form of "electric field energy minus magnetic field 
energy". The interaction Lagrangian $L_{int}$ couples the particle 
variables to the field variables. The part of the Lagrangian that involves 
the vector potential can be written as the integral over a Lagrangian 
density $\cal L$,
\begin{equation}
{\cal L}\,=\, \frac{1}{8\pi}\,\bigl[\frac{\dot{\bA} 
^2}{c^2}\,- \,(\mbox{curl}\,{\bA})^2\bigr]\,+\,
\frac{1}{c}\,{\bJ}^{\bot}(\br) \cdot {\bA}({\br})
\end{equation}
Since the Coulomb gauge condition is being used as a constraint, the 
transverse current density $\bJ^{\bot}$ has been substituted for the total 
current density $\bJ$ without affecting the Lagrangian.

To verify that $L$ given by Equation (2.1) is indeed a suitable 
Lagrangian, we write down the Euler-Lagrange equations, beginning with 
those for the particle coordinates. 
\begin{eqnarray}
\frac{d}{dt}\,\frac{\partial L}{\partial \dot{q}_{\alpha i}}\,-\,
\frac{\partial L}{\partial q_{\alpha i}}\,=\,0. \nonumber 
\end{eqnarray}
In this particular case we have 
\begin{eqnarray} 
m_\alpha\,\ddot{q}_{\alpha i}\,-\,e_\alpha \Bigl[ \sum_{\beta \ne 
\alpha} \frac{e_\beta}{|{\bq}_\alpha\,-\,{\bq}_\beta|^3}\,(q_{\alpha i} 
\,-\,q_{\beta i}) \,-\,\frac{1}{c}\,\dot{a}_i \Bigr]\,
-\,\frac{e_{\alpha}}{c}\dot{q}_{\alpha j}\Bigl(\frac{\partial 
a_j}{\partial q_{\alpha i}}\,-\,\frac{\partial a_i}{\partial q_{\alpha j}}
\Bigr)\,=\,0. 
\end{eqnarray}
Using the expressions
$ {\bB}\,=\,\mbox{curl}{\bA}$
and
$ {\bE}^{\parallel}\,=\,- \mbox{grad}\phi$, \quad ${\bE}^{\bot}\,=\,-
\frac{1}{c}\dot{\bA}$
with 
$\phi({\br},\,t)\,=\,\sum_{\alpha}\frac{e_\alpha}{|{\bq}_\alpha\,-\,{\br}|}$ 
for the transverse fields in terms of the 
vector potentials, we find that \begin{equation} 
m_\alpha\,\ddot{\bq}_\alpha\,=\,e_\alpha\,{\bE}({\bq}_\alpha,\,t)\,+\,
\frac{e_\alpha}{c}\dot{\bq}_\alpha \times {\bB}({\bq}_\alpha,\,t)
\end{equation}
which is nothing but the second law of Newton with the Lorentz force.  

To obtain the field equations, the functional derivatives of $L$ must 
first be calculated from the Lagrangian density $\cal L$. The 
Euler--Lagrange field equations
$$\frac{\partial}{\partial t}\, \frac{\delta L}{\delta \dot A_{i}}\, -\,
\frac{\delta L}{\delta A_i}\,=\,0$$
in this case give
$$
\mbox{curl}\,{\bB}\,=\,\frac{1}{c}\,\dot{\bE}\,+\,\frac{4\pi}{c}\,{\bJ}^{\bot}
$$
which leads to the usual evolution equation for the vector 
potential in the Coulomb gauge:
\begin{equation}
\mbox{curl\,curl}\bA\,+\,\frac{1}{c^2}\,\ddot{\bA}\,=\,\frac{4\pi}{c}
\bJ^{\bot}.
\end{equation}
The conjugate momenta corresponding to the Lagrangian (2.1) are defined in 
the usual way as
\begin{equation}
{\bp}_\alpha\,=\,\frac{\partial L}{\partial \dot{\bq}_\alpha} \,=\,
m\,\dot{\bq}_\alpha\,+\,\frac{e_\alpha}{c}\,{\bA}({\bq},\,t)
\end{equation}
and
\begin{equation}
{\bpi}({\br})\,=\,\frac{\partial L}{\partial \dot{\bA}({\br})}\,=\,
\frac{\dot{\bA}({\br})}{4\pi\,c^2}\,(=\,-{\bE}({\br})/4\pi\,c).
\end{equation}
Proceeding in the conventional way we get 
\begin{eqnarray}
H[{\bpi}, {\bA}; p,q] \,&=&\, \sum_{\alpha}\,{\bp}_\alpha\cdot 
\dot{\bq}_\alpha \,+\,\int\,{\bpi}\cdot \dot{\bA} dV\,-\, L \nonumber \\
&=& \sum_{\alpha}\frac{1}{2m_\alpha} 
\bigl[{\bp}_\alpha\,-\,\frac{e_\alpha}{c}\,{\bA}({\bq},\,t) \bigr]^2\,+\,
\frac{1}{2}\,\sum_{\alpha \ne \beta}\frac{e_{\alpha}\,e_{\beta}}
{|{\bq}_{\alpha}\,-\,{\bq}_{\beta}|} \nonumber \\
&+& \frac{1}{8\pi}\, \int\, \bigl[(4\pi\,c {\bpi})^2 
\,+ \,(\mbox{curl}\,{\bA})^2\bigr]\, dV
\end{eqnarray} 
The Hamiltonian equations reproduce the equations of motions (2.7) and 
(2.8) as they must be. The pair of equations 
$$ {\dot q}_{\alpha i}\,=\,\frac{\partial H}{\partial p_{\alpha i}},
\quad
{\dot A}_{i}\,=\,\frac{\delta H}{\delta \Pi_{i}}$$
are equivalent to the relations between the canonical momenta and 
Lagrangian velocities, where as the equations
$$ {\dot p}_{\alpha i}\,=\,-\,\frac{\partial H}{\partial q_{\alpha i}},
\quad
{\dot \Pi}_{i}\,=\,-\,\frac{\delta H}{\delta A_{i}}$$
yield once again the Lorentz force law (2.7) and the inhomogeneous wave 
equation (2.8) for the vector potential with the transverse current 
density as source.  \vskip 2mm

\subsection{Formation of an equivalent Lagrangian}

\noindent

It is well known that in classical dynamics the addition of a total
time derivative to a Lagrangian leads to a new Lagrangian with the 
equations of motion unaltered. Lagrangians obtained in this manner are 
said to be equivalent. In general, the Hamiltonians following from the 
equivalent Lagrangians are different. Even the relationship between the 
conjugate and the kinetic momenta may be changed.
 
More over, let us notice that the basic equations of any new theory can 
not be introduced strictly deductively. Usually, either they are postulated 
in differential form based on the partial integral conservation laws or 
transformations of basic dynamical variables, whose initial definitions 
usually have some analogs in mechanics. We will mark that we need to do 
so not only by inertia of thinking but also because of the fact that 
most of our measurements have its objects as individual particles or use 
them as test one. The situation is the same in electromagnetism and in 
gravitation.  Attaching of geometrical meaning to any dynamical variables 
plays the crucial role in this case. Even that slight modifications of 
electrodynamic equations that we performed, demands the clear 
understanding of this situation. 

As a simple most example we will consider the "geometrization" of momenta 
in non-relativistic quantum theory when the external classical 
electromagnetic field, acting on free particle, is available. Let us begin 
with the classical Lagrangian:
$${\cal L}(\br, \bv; t)\,=\,\frac{1}{2}\,m {\bv}^2\,-\, q [\phi(\br; t) 
- \bv \cdot \bA(\br;t)],$$
that leads to the equations of motion of particle with charge $q$ and mass 
$m$, moving under the influence of electric and magnetic potentials $\phi$ 
and $\bA$. To move up to Hamiltonian formalism, more closer to the quantum 
one, let us find the canonical momenta of the particle:
$$\bp:=\delta_v\,{\cal L}(\br,\bv;t)=m\bv\,+\,q\bA(\br;t). \eqno(\star)$$
The Hamiltonain in this case is defined as:
$$H:=\bp \bv - {\cal L}\,=\,\frac{1}{2m}[\bp - q\bA]^2 + q\phi. 
\eqno(\star\star)$$
Let us use the initially defined mechanical momenta $\bppi:=m\bv$ 
(gauge invariant quantity that can be directly measured!) to write the 
expression in square bracket in ($\star \star$). We find
$$ [\bp - q\bA]^2 \,=\,[m\bv + q\bA - q\bA]^2 \,=\,\bppi^2.$$
It is of course the simplest tautology, nevertheless under introduction of 
electromagnetic field in Schr$\ddot o$dinger equation it leads to some 
subtlety. Quantum physicists random say or write that the gauge invariant 
introduction of external electromagnetic field in Schr$\ddot o$dinger 
equation is occurred by extended momenta (of particle), implying the 
substitution $\bp \to \bp\,-\,q\bA$. 

In fact the operation to move up from the classical description to the 
quantum one with the introduction of electromagnetic potential we imply 
the following substitutions:  $$\bppi:=m\bv=\nabla \qquad \mbox{to} \qquad 
\bp=\bppi+q\bA \to \nabla$$ and $$H=\frac{1}{2m}[\nabla - q\bA]^2$$.  In 
this the mathematical expectation of meansquare deviation between 
"geomtrized" momenta (operator of spatial displacement of particle) and 
field (potential) angular moment of the particle $q\bA$ are equated to the 
quantum mechanical expectation of evolution operator of the particle 
considered: 
$$i \hbar \Psi^{+}\frac{\partial}{\partial t}\Psi\,=\,
\frac{1}{2m}\Psi^{+}\bigl[(i\hbar \nabla + q\bA)^2 + q\phi\bigr]\Psi. 
\eqno(\star\star\star)$$
In deduction of Schr$\ddot o$dinger equation in [5] the 
similar view is taken into account. Power a.o. [5] (see also [14]) begin 
the introduction of electrical polarization in Schr$\ddot o$dinger 
equation starting with the equivalent Lagrangian.

An equivalent Lagrangian to that of (2.1) is [5] 
\begin{equation} 
L^{new}\,=\,L\,-\,\frac{1}{c}\frac{d}{dt}\bigl[\int\,{\bP}({\br})\cdot 
{\bA}({\br})\,dV\bigr].
\end{equation}  
Taking into account the Coulomb gauge condition $\mbox{div}{\bA}\,=\,0$ 
and toroid contribution we may substitute \begin{equation} 
{\bP}\,\Longrightarrow \,{\bP}^{\bot}\,+\,{\bP}^{\parallel}\,+\, 
\mbox{curl}\,{\bT}^e, \end{equation} with the contribution of 
${\bP}^{\parallel}$ being vanished.

Thus we have the following equivalent Lagrangian 
\begin{equation} 
L^{new}\,=\,L\,-\,\frac{1}{c}\frac{d}{dt}\bigl[\int\,({\bP}^{\bot}({\br})\,
+\, \mbox{curl}{\bT}^e({\br}))\cdot{\bA}({\br})\,dV\bigr].
\end{equation}
Writing it in the explicit form we get
\begin{eqnarray}
L^{new}\,&=&\,L_{par}\,+\,L_{rad}\,+\,
\frac{1}{c}\,\int\limits\,\bigl({\bJ}^{\bot}({\br})\,-\,\dot{\bP} 
^{\bot}({\br})\,-\,\mbox{curl}\dot{\bT}^e({\br})\bigr)\cdot {\bA}({\br})\,dV\,
\,- \nonumber \\ &-&\,\frac{1}{c}
\, \int\,({\bP}^{\bot}({\br})\,+\, \mbox{curl}{\bT}^e({\br}))\cdot 
\dot{\bA}({\br})\,dV\, \end{eqnarray}
The field conjugate to the vector potential $\bA$ is now
\begin{equation}
{\bpi}({\br})\,=\,\frac{\partial L^{new}}{\partial \dot{\bA}({\br})}\,=\, 
-\,\frac{{\bE}^{\bot}({\br})\,+\,4\pi\,({\bP}^{\bot}({\br})\,
+\,\mbox{curl}{\bT}^e({\br}))}{4\pi\,c}
\end{equation} 
In fact, if we define the new vector field in a medium $\cD$ 
by the relation \begin{equation} 
{\cD}({\br})\,=\,{\bE}^{\bot}({\br})\,+\,4\pi\,({\bP}^{\bot}({\br})\,+\, 
\mbox{curl}{\bT}^e ({\br}))\,=\,{\bD}^{\bot}({\br})\,+\,4\pi\,
\mbox{curl}{\bT}^e({\br})\end{equation}
for the conjugate field we get
\begin{equation}
{\bpi}({\br})\,=\,-\,\frac{{\cD}({\br})}{4\pi\,c}
\label{}
\end{equation}
Now from the definition of $\cD$ we get 
\begin{equation}    
\mbox{curl}{\cD}({\br})\,=\,-\,\frac{1}{c}\,\dot{\bB}({\br})\,+\,
4\pi\,(\mbox{curl}{\bP}({\br})\,+\,\mbox{curl\,curl}{\bT}^e({\br})),  
\end{equation}          
under  $\mbox{curl}{\bE}({\br})\,=\,-\,\frac{1}{c}\,\dot{\bB}({\br})$,  
since only the free field $\bE$ is generated due to the change of 
magnetic field $\bB$ as before. 

The new Lagrangian is a function of the variables ${\bq}_\alpha$,
$\dot{\bq}_\alpha$ and a functional of the field variables ${\bA}$, 
$\dot{\bA}$, and the equations of motion follow from the variational 
principle. We have
\begin{equation} \frac{\partial L^{new}}{\partial 
{\bA}({\br})}\,=\,\frac{1}{c}\, \bigl({\bJ}^{\bot}({\br})\,-\,\dot{\bP} 
^{\bot}({\br})\,-\,\mbox{curl}\dot{\bT}^e({\br})\bigr)\,=\,
\frac{1}{c}\,\bigl(c\,\mbox{curl}{\bM}^{\bot}({\br})\,-\,\mbox{curl}\dot{\bT}^e
({\br})\bigr),
\end{equation}
where the relation
$c\,\mbox{curl}{\bM}^{\bot}({\br})\,=\,{\bJ}^{\bot}({\br})\,-\,\dot{\bP} 
^{\bot}({\br})$ is taken into account and
\begin{equation}
\frac{\partial L^{new}}{\partial (\partial 
{\bA}_i({\br})/\partial x_j)}\,=\,\frac{1}{4 \pi}\,\varepsilon_{ijk}\,B_k 
\end{equation} 
as usual.
Applying the Euler-Lagrange equations of motion 
\begin{equation}
\frac{\partial}{\partial t}\,
\frac{\partial L^{new}}{\partial \dot{\bA}_i({\br})}\,+\,\sum_{j}
\frac{\partial}{\partial x_j}\,
\frac{\partial L^{new}}{\partial (\partial
{\bA}_i({\br})/\partial x_j)}\,-\,
\frac{\partial L^{new}}{\partial {\bA}_i({\br})}\,=\,0
\end{equation}
we obtain
\begin{equation}
-\,\frac{\dot 
{\cD}({\br})}{4\pi\,c}\,+\,\frac{1}{4\pi}\,\mbox{curl}{\bB}({\br})\,
-\, \frac{1}{c}\, 
\bigl(c\,\mbox{curl}{\bM}^{\bot}({\br})\,-\,\mbox{curl}\dot{\bT}^e({\br})\bigr)
\,=\,0.
\end{equation}
Introducing the following re-notations, corresponding to the initial 
identification by Maxwell for local conastrain between the derivatives of 
electric and magnetic fields: \begin{equation} 
\frac{1}{c}\,\dot{\bT}^{e,m}\mid_{\Omega}\,\Longrightarrow\,\mp\,\mbox{curl}\,
{\bT}^{m,e}\mid_{\Omega} \end{equation}
we may rewrite (2.23) as
\begin{equation}
\mbox{curl}{\bB}({\br})\,=\,\frac{1}{c}\dot{\cD}({\br})\,+\,4\pi
\bigl(\mbox{curl}{\bM}^{\bot}({\br})\,+\,\mbox{curl\,curl}{\bT}^m({\br})\bigr)
\end{equation}
The relation (2.24) demands some comments. Both $\bT^e$ and $\bT^m$ 
represent themselves in essence closed isolated lines of electric and 
magnetic fields. So they have to obey the usual differential relations 
similar to the free Maxwell equations (see in [2] and [6]). However, 
remark that signs in (2.24) are inverse in comparison with the 
corresponding Maxwell equations because the direction of electric dipole 
is accepted to be chosen opposite to its inner electric field [7].

If we 
define the auxiliary field $\bH$ to be \begin{equation} 
{\bH}({\br})\,=\,{\bB}({\br})\,-\,4\pi\, 
({\bM}^{\bot}({\br})\,+\,\mbox{curl}{\bT}^m({\br})), \end{equation} then 
it deduces to 
$$ \mbox{curl}{\bH}\,=\,\frac{1}{c}\dot{\cD}.$$  
But the latter formula is unsatisfactory from the physical point of view.  
It is easy to image the situation when $\bB$ and $\bM^{\bot}$ are absent, 
because the medium may be composed from isolated aligned dipoles $\bT^m$ 
[1,\,2]  and each $\bT^m$ is the source of free-field 
(transverse-longitudinal) potential but not $\bB$ [3]. So the transition 
to the description by means of potentials is inevitable.
\vskip 2mm
\subsection{Hamiltonian for new Lagrangian}

\noindent

Proceeding as before we get 
\begin{eqnarray}
H^{new}[{\bpi}, {\bA}; p,q] \,&=&\, \sum_{\alpha}\,{\bp}_\alpha\cdot 
\dot{\bq}_\alpha \,+\,\int\,{\bpi}\cdot \dot{\bA} dV\,-\, L^{new} 
\nonumber \\ &=& \sum_{\alpha}\frac{1}{2m_\alpha} 
\bigl[{\bp}_\alpha\,-\,\frac{e_\alpha}{c}\,{\bA}({\bq},\,t) \bigr]^2\,+\,
\frac{1}{2}\,\sum_{\alpha \ne \beta}\frac{e_{\alpha}\,e_{\beta}}
{|{\bq}_{\alpha}\,-\,{\bq}_{\beta}|} \nonumber \\
&+& \frac{1}{8\pi}\, \int\, \bigl\{[4\pi(\bP + \mbox{curl}\,\bT^e)-\cD]^2 
\,+ \,(\mbox{curl}\,{\bA})^2\bigr]\, dV  \nonumber \\
&+&\frac{1}{c}\,\int\limits\,\bigl(\dot{\bP} 
^{\bot}\,+\,\mbox{curl}\dot{\bT}^e\bigr)\cdot {\bA}\,dV.
\end{eqnarray} 
Taking into account $ \frac{1}{c} \dot{\bP}^{\bot} = \frac{1}{c}
{\bJ}^{\bot}\,-\,\mbox{curl}{\bM}^{\bot} $ and the relation (2.24) and
also $\mbox{div}[\bM \times \bA]\,=\,\bA\cdot\mbox{curl}\,\bM\,+\,
\bM\cdot\mbox{curl}\,\bA$ the fourth term of the foregoing equality can be 
reconstructed after integration and $H$ can be presented as
\begin{eqnarray}
H^{new}[{\bpi}, {\bA}; p,q] \,&=&\,
\sum_{\alpha}\frac{1}{2m_\alpha} 
\bigl[{\bp}_\alpha\,-\,\frac{e_\alpha}{c}\,{\bA}({\bq},\,t) \bigr]^2\,+\,
\frac{1}{2}\,\sum_{\alpha \ne \beta}\frac{e_{\alpha}\,e_{\beta}}
{|{\bq}_{\alpha}\,-\,{\bq}_{\beta}|} \nonumber \\
&+& \frac{1}{8\pi}\, \int\, \bigl\{[4\pi(\bP + \mbox{curl}\,\bT^e)-\cD]^2 
\,+ \,(\mbox{curl}\,{\bA})^2\bigr\}\, dV  \nonumber \\
&+&\frac{1}{c}\,\int\limits\,\bJ^{\bot}\cdot \bA\,dV\,-\,
\int\limits 
{\bM}\cdot{\bB}\,dV\,-\,\int\limits{\bB}\cdot\mbox{curl}\,{\bT}^m\,dV.  
\end{eqnarray} 
The Hamiltonian equations in this case give 
$$ {\dot q}_{\alpha 
i}\,=\,\frac{\partial H}{\partial p_{\alpha i}}\,=\, 
\sum_{\alpha}\frac{1}{m_{\alpha i}}\bigl(p_{\alpha i} - 
\frac{e_{\alpha i}}{c}A_i(\bq_\alpha,t)\bigr), \quad {\dot 
A}_{i}\,=\,\frac{\delta H}{\delta \Pi_{i}}\,=\,
c[4\pi(\bP+\mbox{curl}\,\bT^e)-\cD]_i $$ 
which together with
$$ {\dot \Pi}_{i}=-\,\frac{\delta H}{\delta A_{i}}=
\sum_{\alpha}\frac{e_\alpha}{c\,m_{\alpha i}}\bigl(p_{\alpha i} - 
\frac{e_{\alpha i}}{c}A_i(\bq_\alpha,t)\bigr) -\frac{1}{c}(\dot{\bP}+ 
\mbox{curl}\,\dot{\bT}^e)_i 
-\frac{1}{4\pi}[\nabla\times(\nabla\times\bA)]_i,$$ 
$${\dot p}_{\alpha i}=-\,\frac{\partial H}{\partial q_{\alpha i}}=
\sum_{\alpha}\frac{e_\alpha}{c\,m_{\alpha i}}\bigl(p_{\alpha i} - 
\frac{e_{\alpha i}}{c}A_i(\bq_\alpha,t)\bigr) \Bigl(
\frac{\partial A_j}{\partial q_{\alpha i}}-
\frac{\partial A_i}{\partial q_{\alpha j}}\Bigr)-
\frac{1}{2}\sum_{\alpha}\frac{e_\alpha\,e_\beta}{|\bq_\alpha-\bq_\beta|}
(q_\alpha-q_\beta)
$$ 
once again yield the Lorentz force law (2.7) and the inhomogeneous wave 
equation (2.8) for the vector potential.

\vskip 2mm
\subsection{Canonical quantization}

\noindent
Let us now draw the Schr$\ddot o$dinger picture operators corresponding to 
the quantities obtained above. To do so we need to interpret the 
coordinates $\bq_\alpha$ of the particles together with their conjugate 
momenta $\bp_\alpha$ and the {\it coordinates} $\bA(\br)$ of the field 
together with their conjugate momenta $\bpi(\br)$ as operators in Hilbert 
space. Thus we have six operators $q_{\alpha i}$ and $p_{\alpha i}$, 
$(i=1,2,3)$ associated with each particle $\alpha$ and six operators 
$A_i(\br)$ and $\Pi_i(\br)$, $(i=1,2,3)$ associated with each field point 
$\br$. These operators are Schr$\ddot o$dinger picture operators since 
they are time-independent. The corresponding Heisenberg picture operators 
may be written as $\bq_\alpha(t),\,\bp_\alpha(t),\,\bA(\br,\,t),\, 
\bpi(\br,\,t)$ respectively.

The Schr$\ddot o$dinger operators are to satisfy the following canonical 
commutation relations, precisely
\begin{eqnarray}
 [q_{\alpha i}, q_{\beta j}] & = & 0, \quad
[p_{\alpha i}, p_{\beta j}] \,=\, 0   \nonumber \\%
{[q_{\alpha i}, p_{\beta j}]} & = & i \hbar\, \delta_{\alpha 
\beta}\,\delta{ij} 
\end{eqnarray}
for the particle variables and
\begin{eqnarray}
[A_i(\br), A_j(\br^\prime)]&=& 0, \quad
[\Pi_i(\br), \Pi_j(\br^\prime)]\,=\,0  \nonumber \\%
{[A_i(\br), \Pi_j(\br^\prime)]}&=& i \hbar\,\delta_{ij}^{\bot}(\br - 
\br^\prime) \end{eqnarray} for the field variables. These equations are to 
hold for all $\alpha$ and $\beta$ and all $i$ and $j$. The commutator 
bracket of any particle variable with any field variable is to vanish. The 
canonical commutation relations are also satisfied by the Heisenberg 
dynamical variables, provided all operators are evaluated at the same time 
$t$. These follows from the equations (2.29) and (2.30) and the relation 
between two pictures:  \begin{equation} \Omega 
(t)\,=\,e^{(i/\hbar)\,Ht}\,\Omega e^{-(i/\hbar)\,Ht} \end{equation} where 
$H$ is the hamiltonian operator and $\Omega (t)$ is the Heisenberg 
operator corresponding to the Schr$\ddot o$dinger operator $\Omega$. The 
operator $\Omega (t)$ satisfies the Heisenberg equation
\begin{equation}
\dot{\Omega}(t)\,=\,\frac{1}{i \hbar}\,[\Omega (t), H],
\end{equation}
while the Schr$\ddot o$dinger picture operator corresponding to 
$\dot{\Omega}(t)$ is given by
\begin{equation}
\dot{\Omega}\,=\,\frac{1}{i \hbar}\,[\Omega, H].
\end{equation}
Taking the expression (2.28) for the Hamiltonian one can easily verify 
that the equation (2.32) together with the commutation relations leads to 
the expected equations of motion. 

\vskip 2mm
\subsection{Formal deduction of equations of electromagnetotoroidics.}

\noindent

Let us suppose that in an electromagnetic medium there are no free charges
and currents. So we may rewrite usual Maxwell equation in the following 
symmetrical form:
\begin{eqnarray}
\mbox{curl}\,\bB\,=\,\frac{1}{c} \dot{\bD}\,+\,4\pi\,\mbox{curl}\,\bM, 
\quad \bB\,=\,\bH\,+\,4\pi\bM \,\,\mbox{in the whole}\,\, \Re^3,
\nonumber \\
\mbox{curl}\,\bD\,=\,-\frac{1}{c} \dot{\bB}\,+\,4\pi\,\mbox{curl}\,\bP, 
\quad \bD\,=\,\bE\,+\,4\pi\bP\,\, \mbox{in the whole}\,\, \Re^3,
\end{eqnarray}
being of these equations interchange to each other as before through the 
self-reciprocal exchanges $\bB \to \mp \bD, \quad \bD \to \pm \bB, 
\quad \bM \to \mp \bP, \quad \bP \to \pm \bM$ and the conditions 
$\mbox{div}\,\bB\,=\,0$ and $\mbox{div}\,\bD\,=\,0$ are fulfilled. 
Hence is valid the following changes [8] 
\begin{eqnarray}
\bB \Longrightarrow 
\mbox{curl}\,\bal^m\,+\,\dot{\bal}^e, \qquad \bD \Longrightarrow 
\mbox{curl}\,\bal^e\,-\,\dot{\bal}^m, \nonumber \\
\mbox{curl}\,\bM 
\Longrightarrow \mbox{curl}\,\bM\,+\,\mbox{curl\,curl}\,\bT^m \qquad 
\mbox{curl}\,\bP \Longrightarrow \mbox{curl}\,\bP\,+\, 
\mbox{curl\,curl}\,\bT^e 
\end{eqnarray}
whereas the last substitution has already been implied canonically in the 
fore going paragraphs. As a result we obtain for basic equations 
\begin{eqnarray} \mbox{curl\,curl}\,{\bal}^m\,+\,\ddot{\bal}^m\,=\,4\pi\, 
(\mbox{curl}\,\bM\,+\,\mbox{curl\,curl}\,\bT^m), \nonumber \\
\mbox{curl\,curl}\,{\bal}^e\,+\,\ddot{\bal}^e\,=\,4\pi\,
(\mbox{curl}\,\bP\,+\,\mbox{curl\,curl}\,\bT^e), 
\end{eqnarray} 
It is necessary to emphasize that the potential descriptions 
electrotoroidic and magnetotoroidic media are completely separated. The 
properties of the magnetic and electric potentials $\bal^m$ and $\bal^e$ 
under the temporal and spatial inversions are opposite [2]. We 
also see that the substitutions (2.28) in (2.25) produce (2.34) as well.

If $\mbox{div}\bD \ne 0$ and in the medium there does exist free current 
we have to remember Dirac's approach to the constrained systems and use 
along for instance [9] its application to electrodynamics of continuous 
media.  
\vskip 2mm
\subsection{Solution to
the electromagnetotoroidic equations}
\noindent

Let us first solve the
static equations. In this case we write
\begin{equation}
\mbox{curl\,curl}({\bal}^m\,-\,4\pi{\bT}^m)\,=:\,\mbox{curl\,curl}{\ba}^m\,=\,
4\pi\,\mbox{curl}{\bM}^{\bot}({\br}),
\end{equation}
\begin{equation}
\mbox{curl\,curl}({\bal}^e\,-\,4\pi{\bT}^e)\,=:\,\mbox{curl\,curl}{\ba}^e\,=\,
4\pi\,\mbox{curl}{\bP}^{\bot}({\br}),
\end{equation}
here $\ba$ is an analog of $\bH$. From the vector analysis we can see
\begin{equation}
\mbox{curl\,curl}{\ba}^m\,=\,\mbox{grad}\,\mbox{ 
div}{\ba}^m\,-\,\nabla^2{\ba}^m
\,=\,- 4\pi\,\mbox{grad}\,\mbox{div}{\bT}^m\,-\,\nabla^2{\ba}^m
\end{equation}
where we use the condition $\mbox{div}{\bal}^m\,=\,0$ The solutions of the 
equations (2.37) and (2.38) can be written as  
\begin{equation} 
{\ba}^m\,=\,\int\,\frac{\mbox{curl}^{\prime}{\bM}({\br}^{\prime})}{|{\br}\,-\,
{\br}^{\prime}|} \,dV^{\prime}
\,+\,\nabla\,\int\,\frac{\mbox{div}^{\prime}{\bT}^m({\br}^{\prime})}
{|{\br}\,-\,{\br}^{\prime}|} 
\,dV^{\prime} 
\end{equation}
Analogically we get
\begin{equation}
{\ba}^e\,=\,\int\,\frac{\mbox{curl}^{\prime}{\bP}({\br}^{\prime})}
{|{\br}\,-\,{\br}^{\prime}|} \,dV^{\prime}
\,+\,\nabla\,\int\,\frac{\mbox{div}^{\prime}{\bT}^e({\br}^{\prime})}
{|{\br}\,-\,{\br}^{\prime}|} 
\,dV^{\prime} 
\end{equation}
Now let us return to the general case. Adding 
$-\,(4\pi/c^2)\,\ddot{\bT}^{m,e}$ in the both sides of the equations 
(2.35) and (2.36) for $\ba$ we get the wave equations 
\begin{equation} 
\mbox{curl\,curl}{\ba}^m\,+\,\frac{1}{c^2}\,\ddot{\ba}^m \,=\, 4\pi\, 
\mbox{curl}{\bM}^{\bot}({\br})\,-\,\frac{4\pi}{c^2}\,\ddot{\bT}^m({\br})
\end{equation}
and
\begin{equation}
\mbox{curl\,curl}{\ba}^e\,+\,\frac{1}{c^2}\,\ddot{\ba}^e\,=\, 4\pi\,
\mbox{curl}{\bP}^{\bot}({\br})\,-\,\frac{4\pi}{c^2}\,\ddot{\bT}^e({\br})
\end{equation}
The solutions to these equations can be written as [10]
\begin{equation} {\ba}^m\,=\,-\frac{1}{4\pi}\,\int\limits_{all\,space} 
\frac{\bigl[-\nabla^{\prime}(\nabla^{\prime}\cdot {\bT}^{m\,\prime})\,-\,
4\pi\,\nabla^{\prime}\times 
{\bM}^{\prime})\,+\,(4\pi/c^2)\ddot{\bT}^{m\,\prime}\bigr]} 
{|{\br}\,-\,{\br}^{\prime}|}\,dV^{\prime}
\end{equation}
\begin{equation}
{\ba}^e\,=\,-\frac{1}{4\pi}\,\int\limits_{all\,space}
\frac{\bigl[-\nabla^{\prime}(\nabla^{\prime}\cdot {\bT}^{e\,\prime}))\,-\,
4\pi\,\nabla^{\prime}\times 
{\bP}^{\prime})\,+\,(4\pi/c^2)\ddot{\bT}^{e\,\prime}\bigr]} 
{|{\br}\,-\,{\br}^{\prime}|}\,dV^{\prime}
\end{equation}
here $\bM^{\prime}$ denotes $\bM(\br^{\prime})$. Under the Coulomb gauge 
the substitution $\mbox{div}\,\ba^{m,e}\,=\,-4\pi\,\mbox{div}\,\bT^{m,e}$ 
is valid. The brackets [\,] in the above equations are the "retardation 
symbol".  This symbol indicates a special space and time dependence of the 
quantities to which it is applied and is defined by the identity 
$$[f]\,\equiv\,f(x^{\prime},y^{\prime},z^{\prime},t-r/c).$$

It is obvious from the foregoing solutions that the quantities that 
an experimentator measures are $\bal$ and $\ba$ and the 
fields, they generate, differ from the old ones. It leads to the fact that 
we no longer work with $\bB$ and $\bD$ but with some  
new quantities which need to be defined differently. So instead of (2.35) 
we should now write  
\begin{equation}
\bbe\,=\,\mbox{curl}\,\bal^m\,+\,\dot{\bal}^e, \qquad 
\bde\,=\,\mbox{curl}\,\bal^e\,-\,\dot{\bal}^m.  
\end{equation}
Note that the quantities $\bbe$ and $\bde$ may be measured and possess 
physical significance. For example, the experimentator, measuring magnetic 
field in a needle-like hole, dug in magnetic, not along the principal 
axis of magnetization, beside the contribution of Ampere current of 
magnetization $\mbox{curl}\bM$ measures the contribution of inhomogeneous
toroidazation that is obvious from (2.44) and (2.45). Thus  
introduction of potential and magnetic field based on the magnetization of 
media only, generally speaking is incorrect [3].

\vskip 1cm
\section{Multipolar interactions in Schr$\ddot o$dinger equations} 
\setcounter{equation}{0}
\vskip 5mm
\noindent
A large number of works have been devoted to this problem. Unfortunately 
most of them are confusing. We demonstrate it on the basis 
of detail paper by K.\ Haller and R.\ B.\ Sohn [11]. The 
matter is that they begin with the expression independent of the 
scalar-longitudinal contributions $(\rho, \mbox{div}\bj)$ and after the 
integrations by parts  in the final expression restore them. It is a very 
common methodical error.

\vskip 2mm
\subsection{Description of non-relativistic
interaction of electrons and photons} 
\noindent
Here we proceed and set the problem in terms used 
in [11].  The Hamiltonian that describes the interaction between photons 
and non-relativistic Schr$\ddot o$dinger electrons is given by 

\begin{eqnarray}
H_c&=&H_0\,-\,\int\,{\bJ}({\br})\cdot
{\bA}^T({\br})\,d{\br}\,+\,\frac{e}{2m} 
\int\,\rho({\br}){\bA}^T({\br})\cdot {\bA}^T({\br})\,d{\br}\,+ \nonumber 
\\&+&\,\int\,\frac{\rho({\br})\rho({\br'})}{8\pi 
|{\br}\,-\,{\br'}|}\,d{\br}\,d{\br'}, 
\end{eqnarray}

where $\bA^T$ is the transverse vector potential with $\mbox{div}\,
\bA^T\,=\,0$. $H_0$ is the Hamiltonian for noninteracting electrons and 
photons that can be represented by
\begin{equation}
H_0\,=\,H_0(e)\,+\,H_0(\gamma),
\end{equation}
where 
\begin{equation}
H_0(e)\,=\,\int\,\psi^+(\br)[-(2m)^{-1}\,\nabla^2\,+\,V(\br)]\psi(\br)d\br
\end{equation}
with $V(\br)$ representing an external short-range potential (for example, 
the shielded Coulomb potential of a static nucleus). $H_0(\gamma)$ is the 
Hamiltonian for free transverse photons and is given by
\begin{equation}
H_0(\gamma)\,=\,\frac{1}{2}\int\,[\bE^T(\br)^2\,+\,\bB(\br)^2]d\br,
\end{equation}
where $\bE^T(\br)$ and $\bB(\br)$ represent the transverse electric and 
magnetic field, respectively. In equation (3.1) $\rho(\br)$ represents the 
charge density $\rho(\br)\,=\,e\,\psi^+(\br)\psi(\br)$ and $\bJ(\br)$ 
represents a current
\begin{equation}
\bJ(\br)\,=\,\frac{ie}{2m}[\nabla 
\psi^+(\br)\,\psi(\br)\,-\,\psi^+(\br)\,\nabla\psi(\br)].
\end{equation}
The commutation rules of these fields are 
\begin{equation}
\{\psi^+(\br),\psi(\br')\}\,=\,\delta(\br\,-\,\br')
\end{equation}
for the electron field; for the photon field, $-\bA^T(\br)$ has the 
transverse electric field $\bE^T(\br)$ as its conjugate momentum, so that 
the nonlocal commutation rule for the transverse components is given by
\begin{equation}
[A_{i}^{T}(\br),E_{j}^{T}(\br')]\,=\,-\bigl(\delta_{i,j}\,\delta(\br\,-\,\br')
\,+\,\frac{\partial}{\partial r_i}\, \frac{\partial}{\partial r_j}\,
\frac{1}{4\pi |\br\,-\,\br'|} \bigr).
\end{equation} 
The current $\bJ(\br)$ is conserved under the time dependence provided by 
the Hamiltonian so that
\begin{equation}
\mbox{div}\,\bJ(\br)\,=\,-i[H_0,\rho(\br)].
\end{equation}
The current
\begin{equation}
\bj(\br)\,=\,\bJ(\br)\,-\,(e/m)\,\rho(\br)\,\bA^T(\br)
\end{equation}
is conserved under the time dependence provided by the Hamiltonian so that
\begin{equation}
\mbox{div}\,\bj(\br)\,=\,-i[H_c,\rho(\br)].
\end{equation}
$H_c$ may be expressed as
\begin{eqnarray}
H_c\,=\,H_0\,-\,\int\,[\bJ(\br)\,+\,\bj(\br)]\cdot 
{\bA}^T({\br})\,d{\br}\,+\, 
\int\,\frac{\rho({\br})\rho({\br'})}{8\pi 
|{\br}\,-\,{\br'}|}\,d{\br}\,d{\br'}. 
\end{eqnarray}
When $\psi$ is quantized in orbitals that are solutions of
\begin{equation}
[-(2m)^{-1}\,\nabla^2\,+\,V(\br)\,-\,\omega_i]\,U_i\,=\,0,
\end{equation}
(where $U_i$ includes both bound and continuum states), $H_0(e)$ is given 
by
\begin{equation}
H_0(e)\,=\,\sum_{n}\,e_{n}^{+}\,e_n\,\omega_n,
\end{equation}
where $e_{n}^{+}$ and $e_n$ designate creation and annihilation operators, 
respectively, for electrons in the $n$ orbitals. $H_0(\gamma)$ can be 
represented as 
\begin{equation}
H_0(\gamma)\,=\,\sum_{i=1,2}\,\int\,d\bk\,|\bk|\ba_{\epsilon(i)}^{+}(\bk)\,
\ba_{\epsilon(i)}(\bk),
\end{equation}
and describes free photons in the two polarization modes.
\vskip 2mm
\subsection{Introduction of toroid moments}
\noindent
A formal procedure was first proposed by E.\ G.\ P.\ Rowe [12] permits to 
obtain the complete solution to the problem of multipole expansion of 
electromagnetic current [3], which consists of the replacing of some 
vector function of current by the three (in general unlimited) series of 
multipole parameters. The multipole parametrization of interaction 
Hamiltonian of an arbitrary system with external fields under the Coulomb 
gauge has the form [3]
\begin{eqnarray}
H_c&=&-\sum_{l=1}^{\infty} \sum_{m=-l}^{l} \sum_{n=0}^{\infty}
\frac{(2l\,+\,1)!!}{2^n\,n!(2l+2n+1)!!}\,\sqrt{\frac{2l+1}{4\pi}} \times
\nonumber \\ &\times&
\{l^{-1}M_{lm}^{(2n)}(t)Y_{lm}(\nabla)\triangle^n(\br\cdot\bB)|_{r=0}+
\\  
&+& l_{-1}[\dot{Q}_{lm}(t)\delta_{n,0}\triangle^{-1}-T_{lm}^{2n}(t)]
Y_{lm}(\nabla)\triangle^n [(1/c)\br\cdot \dot{\bD}+(4\pi/c) \br\cdot 
\bj]_{r=0}\}.  \nonumber
\end{eqnarray}
where $\dot {Q}_{lm}(t)$, connected with Coulomb multipole moments 
of the charge distribution of the system are  \begin{equation} \dot 
Q_{lm}(t)\,=\,\sqrt{4\pi l}\,\int\,r^{l-1}\,Y_{l\,l-1\,m}^{*}(\hat 
r)\,\bj(\br,t)\,d\br, \end{equation} $M_{lm}^{(2n)}(t)$ are the 
magnetic multipole moments or their radii \begin{equation} 
M_{lm}^{(2n)}(t)\,=\,\frac{-i}{c}\,\sqrt{\frac{l}{l+1}}\,
\sqrt{\frac{4\pi}{2l+1}}\,\int\,r^l\,Y_{llm}^{*}(\hat 
r)\,\bj(\br,t)\,d\br, 
\end{equation}
and $T_{lm}^{(2n)}(t)$ are third family of multipole moments [13], the 
toroid moments and their radii, namely 
\begin{equation}
T_{lm}^{(2n)}(t)=\,-\frac{\sqrt{\pi}l}{c(2l+1)}
\,\int r^{l+2n+1}[Y_{l\,l-1\,m}^{*}(\hat r)+
\frac{2\sqrt{l/l+1}}{2l+3}Y_{l\,l+1\,m}^{*}(\hat r)]
\cdot\bj(\br,t)\,d\br, 
\end{equation}
In the case considered by Haller and Sohn [11] we have to substitute into 
(3.15) \begin{equation} \dot Q_{lm}(t)\,\equiv\,\frac{\partial \,Q_{lm} 
}{\partial t}\,=\,i[H_c,Q_{lm}] \end{equation} and 
\begin{equation} \bj\,=\,\bJ\,-\,e\rho\,\bA \end{equation} in the formulas 
(3.15), (3.16), (3.17) and (3.18). In this approach there does not appear 
any longitudinal contributions that are fictitious (mutually cancelled 
out) in the expression, deduced by Haller and Sohn.  The formulas (3.16 - 
3.18) obtained by us give correct multipole parametrization  for the 
interaction energy transverse degrees of freedom of a non-relativistic 
system with radiation field.  
\vskip 2mm
\subsection{Toroid moments in Schr$\ddot o$dinger equation}

\noindent
\begin{minipage}{115mm}
\unitlength=1.00mm
\special{em:linewidth 0.4pt}
\linethickness{0.4pt}
\begin{picture}(80.00,99.33)
\put(50.00,50.00){\vector(1,0){30.00}}
\put(50.00,50.00){\vector(0,1){30.00}}
\put(50.00,50.00){\vector(-2,-1){23.67}}
\put(50.00,50.00){\vector(-3,1){20.00}}
\put(70.67,74.33){\circle{6.67}}
\put(39.33,62.67){\makebox(0,0)[lb]{$\bR$}}
\put(30.00,56.67){\vector(1,4){2.67}}
\put(50.00,50.00){\vector(-1,1){17.33}}
\put(32.67,67.33){\vector(0,0){0.00}}
\put(54.00,30.67){\makebox(0,0)[cc]{{\it Figure 1}}}
\put(36.00,54.00){\makebox(0,0)[rt]{$\bq_i$}}
\put(29.33,53.67){\makebox(0,0)[cc]{$e_i$}}
\put(25.00,61.33){\makebox(0,0)[cc]{$\bq_i-\bR$}}
\bezier{132}(15.00,67.00)(16.67,49.67)(32.33,47.00)
\bezier{124}(15.00,67.33)(17.67,82.33)(33.00,84.33)
\bezier{132}(33.33,84.33)(48.00,82.67)(49.00,64.33)
\bezier{120}(49.00,63.33)(47.33,47.67)(33.00,47.33)
\put(69.00,73.00){\makebox(3.67,2.33)[cc]{$\br$}}
\end{picture}
\end{minipage}
\begin{minipage}{45mm}
Now suppose that we consider some electron involved in the molecular 
structure. We may introduce a coordinate system as in Fig. 1. 
Then it may be shown that the Schr$\ddot o$dinger equation for this 
electron interacting with external sources of electromagnetic fields has 
the following form [14] 
\end{minipage}
\begin{eqnarray}
i \hbar \dot{\phi}(\bq)\,=\,\Bigl[-\frac{\hbar^2}{2m}(\nabla^{(q)})^2\,+\,
V(\bq)\,+\,e^2\,\int\,\frac{\bar \phi (\bq') \phi(\bq')}{|\bq\,-\,\bq'|}\,
d\bq\,-\nonumber \\
-\,\int\,\bP(\br,\,\bq)\cdot \bE^{\bot}(\br)\,d\br\,- \,
\int\,\bM(\br,\,\bq)\cdot \bB^{\bot}(\br)\,d\br\,+\nonumber\\
+\,\frac{1}{2mc^2}
\bigl[\int\,\bn(\br,\,\bq)\times \bB(\br)\,d\br\bigr]^2\Bigr]\phi(\bq),
\end{eqnarray}
with $\bn(\br,\,\bq)\,=\,-\frac{e}{2}(\bq\,-\,\bR)\,\delta(\br\,-\,\bR).$

To get the Schr$\ddot o$dinger equation avobe authors used the following 
substitutions in the Lagrangian for an electron field uncoupled from the 
radiation field: $\nabla$ is replaced by $\nabla + \frac{ie}{\hbar 
c}\bA(\bq)$ and the wave function describing electron field initially is 
transformed according to
$$\psi(\bq)\,=\,e^{-S(\bq)}\,\phi(\bq),$$
where $S(\bq)$ is given by
$$S(\bq)\,=\,\frac{1}{\hbar c}\int\limits \bP(\br,\bq)\cdot\bA^{\bot} 
(\br) d\br.$$

There are two ways to introduce here the toroid contributions. The first 
one is straightforward to use the substitution (2.30) Then we obtain 
\begin{equation}
\int\,\bP\cdot \bE^{\bot}\,d\br\, \Longrightarrow
\int\,\bP\cdot \bE^{\bot}\,d\br\,+\,
\int\,\mbox{curl}\bT^e\cdot \bE^{\bot}\,d\br\,=\, 
\int\,\bP\cdot \bE^{\bot}\,d\br\,+\,
\int\,\bT^e\cdot \mbox{curl}\bE^{\bot}\,d\br.
\end{equation}
Analogically
\begin{equation}
\int\,\bM\cdot \bB\,d\br\,
\Longrightarrow\, \int\,\bM\cdot \bB\,d\br\,+\,
\int\,\bT^m\cdot \mbox{curl}\bB\,d\br.
\end{equation}
More reasonable approach is developed in [2]. It goes back to the classic 
multipolar description of quasimolecular structure [9,\,15]. According to 
it we may use immediately the multipole expansion of the densities 
$\bP(\br,\,\bq)$ and $\bM(\br,\,\bq)$ as follows [2]:
\begin{eqnarray}
W^e\,=\,-\int\,\bd(\br,\,\bq)\cdot \bE^{\bot}(\br)\,d\br\,=
- \bQ \cdot \bE^{\bot}\,-\, {\bT}^e\cdot \mbox{curl}\bE - 
\nonumber \\
- \hat{\bP}^{e}\cdot \mbox{curl\,curl}\bE \,-\,\frac{1}{2}Q_{ij}
(\nabla_i \bE_{j}^{\bot}\,+\,\nabla_j \bE_{i}^{\bot})\,-\,
T_{ij}^{e} \nabla_i(\mbox{curl}\bE^{\bot})_{i} - \cdots
\end{eqnarray}
where $\bQ\,=\,\int\,\bd(\br)\,d\br$ is the conventional total electric 
dipole moment of the system, $\bT^e\,=\,\frac{1}{2}\int\,\br\times 
\bd(\br)\,d\br$ is the axial toroid moment and (see [14] and Appendix) 
\begin{eqnarray} \bP(\br,\,\bq)\,=\, 
-e\Bigl\{(\bq\,-\,\bR)\,-\,\frac{1}{2}(\bq\times\bR)\,- \nonumber \\ 
-\frac{e}{2}\bigl[(q\,-\,R)_i\,(q\,-\,R)_j\,-\,\frac{2}{3}
(q\,-\,R)^2\,\delta_{ij}\bigr]\,+\,\cdots 
\Bigr\}\,\delta(\br\,-\,\bR).  
\end{eqnarray}
Analogically for $\bM(\br,\,\bq).$

\noindent
Let us now define the axial toroid moment. 
From the classical point of view on the {\bf origin} of electric dipoles 
we see that the multipole expansion of the distribution density 
$\bP(\br,t)$ contents generally three kinds of dipole moments (see Formula 
(3.24)).  
Working within the scope of classical framework after Power a. 
o. [5] we see that $\bT^e\,=\,0$ for a separate "atom". Really, all the 
electric dipoles are characterized by its space vectors, with the origins 
lying in the origin of the coordinate system (center mass of the system 
considered) and the endpoints in the electron coordinates. The sum of 
these vectors characterizes the total electric dipole moment of the 
system. To demonostrate the forementioned conclusion we consider 
the following example. 

\noindent
\begin{minipage}{115mm}
\unitlength=1.00mm
\special{em:linewidth 0.4pt}
\linethickness{0.4pt}
\begin{picture}(89.00,74.66)
\put(25.33,27.00){\circle*{2.67}}
\put(42.67,60.33){\circle*{4.00}}
\put(80.00,36.00){\circle*{4.00}}
\put(51.33,37.00){\circle*{1.33}}
\put(42.67,62.33){\vector(0,1){7.33}}
\put(44.33,58.67){\vector(1,-1){5.33}}
\put(42.33,58.33){\vector(0,-1){8.33}}
\put(41.33,61.00){\vector(-4,3){7.33}}
\put(44.33,61.00){\vector(3,2){7.00}}
\put(82.00,35.67){\vector(1,0){7.00}}
\put(80.00,33.67){\vector(0,-1){7.67}}
\put(79.67,38.00){\vector(0,1){7.67}}
\put(78.33,36.33){\vector(-2,1){7.00}}
\put(24.33,26.67){\vector(-1,-1){7.00}}
\put(26.33,26.00){\vector(1,-1){6.33}}
\put(51.33,37.00){\vector(-1,3){7.33}}
\put(51.33,36.67){\vector(1,0){27.33}}
\put(51.33,37.00){\vector(-3,-1){25.67}}
\bezier{176}(25.00,29.00)(22.00,52.33)(41.33,58.00)
\bezier{212}(44.67,59.33)(70.67,62.67)(78.33,37.33)
\bezier{316}(79.00,34.33)(62.33,2.67)(26.33,27.00)
\put(24.33,27.33){\vector(-3,4){6.00}}
\put(78.67,35.33){\vector(-2,-1){8.33}}
\put(48.00,30.67){\makebox(0,0)[cc]{$O$}}
\put(43.67,67.33){\line(1,0){17.33}}
\put(63.00,64.67){\makebox(0,0)[cc]{$\bq_{ij}$}}
\put(53.33,40.67){\makebox(0,0)[cc]{$\bR_j$}}
\put(52.33,9.33){\makebox(0,0)[cc]{{\it Figure 2}}}
\put(33.00,59.67){\makebox(0,0)[cc]{$A$}}
\put(25.33,27.67){\vector(1,3){3.33}}
\put(80.67,35.33){\vector(1,-2){5.00}}
\end{picture}
\end{minipage}
\begin{minipage}{45mm}
Let us have a system containing $N$ atoms where each 
j-atom itself forms a subsystem [see Fig. 2]. Let $"O"$ be the center-mass 
of the system as a whole.  If the system contains only one atom, say 
$"A"$, which on his part consists of one nuclei and $n$ electrons, the 
electric dipole moment of this system relating to the point $"O"$ can be 
written as 
\end{minipage}
\begin{equation}
\bQ\,=\,-e\sum_{i=1}^{n}(\bR-\bq_i), \quad \bd\,=\,\bQ\,\delta(\br-\bR).
\end{equation}
Then  the axial toroid 
moment of the such a system  with respect to its proper center-mass is 
\begin{equation}
\bT^{(d)}\,=\,\frac{1}{2} \int\,\bd \times \br \,d\br\,=\, 
\frac{e}{2}\,\sum_{i=1}^{n}(\bR-\bq_i)\times 
(\bR-\bq_i)\,=\,0, \end{equation} 
as $\br_i\,=\,\bR-\bq_i$ and $\bR$ is fixed.
If the molecule contains $N$ atoms then we define the total toroid 
moment of the molecule $\bT^{(d)}$ as a whole [16] with respect to its 
center-mass by the following formula
\begin{equation}
\bT^{(d)}\,=\,\frac{e}{2}\sum_{j=1}^{N}\sum_{i=1}^{n}(\bR_j-\bq_{ij})
\times \bR_j\,=\frac{e}{2}\sum_{j=1}^{N}\sum_{i=1}^{n}[\bq_{ij}\times 
\bR_j] \,\ne 0,\end{equation}
although the axial toroid moment $\bT_{j}^{(d)}$ for each separate 
j-atom with respect to its proper center-mass is equal to zero as before.

As a particular example where electromagnetic properties of molecules are 
described by the axial toroid moment, we point out the phenomenon of 
"aromagnetism" [17].
There exists substances with a closed chains of atoms like benzol ring. 
In the experiment, the microcrystals of the aromatic series 
substances (antrazen, fenantren etc.), suspended either in water or in 
other liquids, were reoriented by an alternating magnetic field so that 
the modulation of polarized light, propagating through the media given, 
was observed. This reorientations can be easily explained if the aromatic 
molecules are considered to possess axial toroid moment $\bT^{(d)}$ (see 
Fig. 3) as an electromagnetic order parameter. The latter can be clarified 
as follows. All molecular wave functions $\Psi$ of the benzol ring, being 
the main fragment of aromatic molecules, are first classified through the 
irreducible representations of the point group symmetry $D_{6h}$ and then 
the asymmetric representations ($E_{2g}$ and $E_{1u}$) are selected among 
them. When such molecules are packed into a molecular crystal, the 
parallel orientation of their axial toroid moments are preferred and the 
crystal as a whole can possess a macroscopic axial toroid moment [18].  
\vskip 3mm
\unitlength=1mm
\special{em:linewidth 0.4pt}
\linethickness{0.4pt}
\begin{picture}(94.67,96.00)
\put(40.00,50.00){\line(0,1){20.00}}
\put(40.00,70.00){\line(5,3){17.33}}
\put(57.33,80.00){\line(3,-2){15.00}}
\put(72.33,70.00){\line(0,-1){20.00}}
\put(72.33,50.00){\line(-3,-2){15.33}}
\put(57.00,39.33){\line(-5,3){17.00}}
\put(40.00,50.00){\circle*{1.33}}
\put(40.00,70.00){\circle*{1.33}}
\put(56.67,79.67){\circle*{2.00}}
\put(72.33,70.00){\circle*{1.33}}
\put(72.33,50.00){\circle*{1.33}}
\put(56.67,39.67){\circle*{1.33}}
\put(64.67,40.00){\vector(-1,0){17.67}}
\put(43.33,43.00){\vector(-1,2){7.67}}
\put(36.33,63.67){\vector(1,2){6.67}}
\put(48.67,80.00){\vector(1,0){17.67}}
\put(68.67,76.67){\vector(2,-3){8.33}}
\put(75.67,55.67){\vector(-1,-2){6.67}}
\put(72.67,49.33){\line(5,-2){5.33}}
\put(79.67,46.33){\line(3,-1){7.33}}
\put(87.00,44.00){\line(0,0){0.00}}
\put(87.67,43.33){\circle*{2.00}}
\put(56.33,38.33){\line(0,-1){5.33}}
\put(56.33,30.67){\line(0,-1){6.00}}
\put(56.33,22.67){\circle*{2.00}}
\put(39.00,49.00){\line(-5,-3){5.00}}
\put(32.33,45.00){\line(-5,-3){5.00}}
\put(26.33,41.33){\circle*{2.00}}
\put(39.00,70.33){\line(-2,1){3.67}}
\put(34.00,73.00){\line(-5,3){4.67}}
\put(28.00,76.33){\circle*{2.00}}
\put(56.67,81.33){\line(0,1){4.00}}
\put(56.67,87.33){\line(0,1){5.67}}
\put(56.67,95.00){\circle*{2.00}}
\put(73.67,71.00){\line(3,2){4.00}}
\put(78.33,74.67){\line(3,2){4.67}}
\put(84.00,78.67){\circle*{2.00}}
\put(53.67,72.67){\makebox(5.00,3.67)[cc]{$C_1$}}
\put(64.67,65.33){\makebox(5.67,3.33)[cc]{$C_2$}}
\put(64.67,49.33){\makebox(5.67,4.00)[cc]{$C_3$}}
\put(53.67,43.33){\makebox(5.00,4.33)[cc]{$C_4$}}
\put(42.67,49.00){\makebox(5.67,4.67)[cc]{$C_5$}}
\put(42.67,65.00){\makebox(5.67,5.00)[cc]{$C_6$}}
\put(59.00,91.00){\makebox(5.33,4.00)[cc]{$X_1$}}
\put(59.00,81.67){\makebox(5.33,4.00)[cc]{$\bd_1$}}
\put(87.33,76.00){\makebox(7.33,4.33)[cc]{$X_2$}}
\put(76.67,67.00){\makebox(6.33,3.67)[cc]{$\bd_2$}}
\put(86.00,36.00){\makebox(7.00,4.00)[cc]{$X_3$}}
\put(72.00,42.00){\makebox(6.33,4.00)[cc]{$\bd_3$}}
\put(60.67,21.33){\makebox(7.00,4.33)[cc]{$X_4$}}
\put(49.33,34.67){\makebox(6.00,3.67)[cc]{$\bd_4$}}
\put(29.33,37.00){\makebox(6.00,3.67)[cc]{$X_5$}}
\put(30.67,49.33){\makebox(6.33,3.67)[cc]{$\bd_5$}}
\put(26.00,69.00){\makebox(6.00,4.00)[cc]{$X_6$}}
\put(35.33,73.67){\makebox(5.33,4.33)[cc]{$\bd_6$}}
\end{picture}  

\noindent
{\it Figure 3.
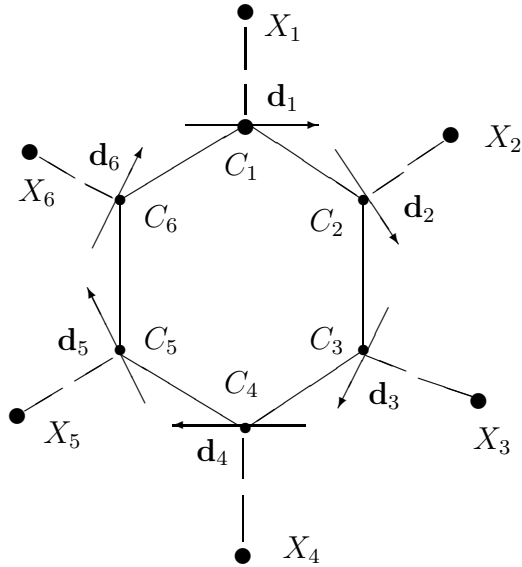
Here is a fragment of an aromatic molecule where $C_i$ is a 
carbon atom, $X_i$ is a hydrogen atom or its substitute with $C_i$. The 
vectors $\bd_i$ denotes the electric dipole moments, being due to the 
molecular orbital with the symmetry $E_{2g}$ of the representation of the 
point group $D_{6h}$. The molecular wave functions $\Psi E_{2g}$ and $\Psi 
E_{1u}$ with this symmetry are built [18].  The molecules in this states 
are shown to have the axial toroid moment.} 
\vskip 5mm \noindent
Now let us see what "aromagnetism" is?
Microscopic   crystals   of   aromatics  like
anthracene,  penthacene,  phenanthrene and others  were  suspended  in
water  or  some  other liquids.  When alternating or rotating magnetic
field was applied to the  suspension  the  change  of  orientation  of
microcrystals  was observed.  This new magnetic property of 
aromatic substances was named "aromagnetism" [17]. This phenomena can 
not be explained by  the standard way like ferromagnetism since the 
organic molecules do not possess magnetic moments neither orbital nor spin 
origin.

The theory of aromagnetism origin was proposed in [18].  It was assumed
there  that  the  sample  of  aromagnetic substance is possessed of an
axial toroid moment $\bG$ (ATM). This moment describes
the     vortex     distribution     of     the    electrical    dipoles
$\bd_a$ deposited  at  points  $\br_a$
and equal to
$$ \bG= (1/2)
\sum [\br_a\,\bd_a], $$
where the sum is taken over the whole sample. Electrodynamics theory of
ATM was developed in [19]. The energy of ATM in external field is equal
to
$$ {\cE}\,=\,-\bG\mbox{curl}\bE= \, \bG\frac{\partial \bH}{\partial ct}.$$ 
Thus toroid  moment  
$\bG$, electrical by its nature, can interact with 
alternating magnetic field $\bH(t)$.  In  the  
following we  consider  as 'aromagnetic'  any  molecule  or macroscopic 
sample that possess toroid moment  $\bG$.  The  main  
feature  that singles  out aromagnetics  among  the other magnetics is 
their interaction with magnetic  field  time  derivative 
$\partial \bH/\partial t$ (not with magnetic field $\bH$ 
itself). The theory previously given  in  [18] may be extended so that 
new  substances  with aromagnetic properties can be predicted. 

Indeed it was  shown  (see [18]) that the cause of molecular aromagnetism 
is the existence of  quantum  molecular  electronic  states  (MO)  where  
the quantum   mean  value  of  toroid  moment  $<\bG>$  of 
molecule is not equal to  zero.  In  [18]  only  molecules  with  group
symmetry $D_{6h}$ were considered. It is obviously that molecules with
the  other  symmetry  groups  may  be  examined.  The  scheme  of  the
examination is the following.  Toroid moment is a pseudovector and for
any given group we know the irreps of  its  transformation  under  the
group  operations.  For  example  in  the group $D_{4h}$ the irreps of
pseudovector are~:  $A_{2g}$ for $G_z$ and $E_g$ for $G_x$  and  $G_y$
components  of  $\bG$.  Considering direct products of
this irreps ($A_{2g}$ and $E_g$ in a given case) with the other irreps
of  this group it is possible to find out the symmetry of MO where the
mean value of toroid moment is not zero  ($E_g$  or  $E_u$  for  group
$D_{4h}$).  Then it is possible to find these MO in explicit form and
calculate the mean value $<\bG>$,  as it was described
in [18].  Toroid  polarization coefficients of the molecule can also be
calculated in analogous way.
\vskip 5mm
\noindent
\begin{minipage}{8cm}
\hspace*{20mm}
\unitlength=0.70mm
\special{em:linewidth 1.0pt}
\linethickness{1.0pt}
\begin{picture}(37.00,37.89)
\put(21.00,22.00){\circle{10.00}}
\emline{18.00}{28.00}{1}{24.00}{28.00}{2}
\emline{24.00}{28.00}{3}{28.00}{22.00}{4}
\emline{28.00}{22.00}{5}{24.00}{16.00}{6}
\emline{24.00}{16.00}{7}{18.00}{16.00}{8}
\emline{18.00}{16.00}{9}{14.00}{22.00}{10}
\emline{14.00}{22.00}{11}{18.00}{28.00}{12}
\emline{24.00}{28.00}{13}{27.00}{34.00}{14}
\emline{27.00}{34.00}{15}{33.00}{34.00}{16}
\emline{24.00}{16.00}{17}{27.00}{10.00}{18}
\emline{27.00}{10.00}{19}{24.00}{5.00}{20}
\emline{14.00}{22.00}{21}{8.00}{22.00}{22}
\emline{4.00}{28.00}{23}{8.00}{22.00}{24}
\emline{18.00}{28.00}{25}{15.00}{34.00}{26}
\emline{28.00}{22.00}{27}{34.00}{22.00}{28}
\emline{18.00}{16.00}{29}{15.00}{10.00}{30}
\put(11.00,27.00){\vector(-1,-2){4.67}}
\put(21.00,10.00){\vector(1,0){11.00}}
\put(31.00,29.00){\vector(-3,4){6.67}}
\put(14.00,37.00){\makebox(0,0)[cc]{\small H}}
\put(2.00,31.00){\makebox(0,0)[cc]{\small H}}
\put(13.00,7.00){\makebox(0,0)[cc]{\small H}}
\put(22.00,2.00){\makebox(0,0)[cc]{\small H}}
\put(37.00,22.00){\makebox(0,0)[cc]{\small H}}
\put(36.00,34.00){\makebox(0,0)[cc]{\small H}}
\put(4.00,22.00){\makebox(0,0)[cc]{\small O}}
\put(29.00,37.00){\makebox(0,0)[cc]{\small O}}
\put(29.00,6.00){\makebox(0,0)[cc]{\small O}}
\end{picture}
\end{minipage}
\begin{minipage}{8cm}
{\it
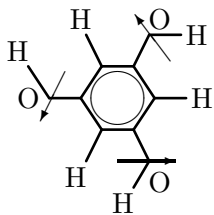
Figure 4. Distribution of dipole moments of oxygen atoms of the molecule
phloroglycine.  It can be estimated from this figure that molecule has
axial toroid moment.}
\end{minipage}
\vskip 5mm
The other important case  of  molecular  aromagnetism  is  the  vortex
distribution  of  dipole moments of oxygen or nitrogen atoms they have
due to their lone pairs of electrons.  On the Figure~4 the stereomer  of
the molecule of phloroglycine is represented and dipole moments of the
oxygen  atoms  are  shown.  The  stability  of  this   stereomer   was
demonstrated  by  numerical  calculation by standard molecular dynamic
method.  It follows from the picture that vortex distribution  of  the
electrical  dipole  moments  on  the oxygen atoms exists and that this
stereomer of phloroglycine is aromagnetic one.  The other examples  of
organic  molecules with the same kind of aromagnetism are given here.

To explain structural aromagnetism let us consider  molecular  crystal
of indandion-1,3  pyridine-betaine  (IPB)  polar  intramolecular salt
(Figure~5,  left).  Due to charge transfer this molecule has  electrical
dipole moment.  The fundamental unit of the crystal of IPB is shown in
Figure~5 (to draw the Figure~5 the picture 1.9 from the book [20] is 
used) and on the  figure  we show  the  dipole  moments  of the molecules.  
It can be seen from the figure that toroid moment of the crystal cell is 
not equal  to  zero, i.e.  this  substance  has  aromagnetic  properties 
which we call structural aromagnetism. 
\vskip 5mm 
\noindent 
\unitlength=0.70mm \special{em:linewidth 1.0pt}
\linethickness{1.0pt}
\begin{picture}(219.00,59.00)
\put(73.00,29.00){\circle*{1.00}}
\put(74.00,39.00){\circle*{1.00}}
\put(77.00,32.00){\circle*{2.00}}
\put(78.00,38.00){\circle*{2.00}}
\put(82.00,25.00){\circle*{1.00}}
\put(82.00,30.00){\circle*{2.00}}
\put(85.00,43.00){\circle*{2.00}}
\put(85.00,48.00){\circle*{1.00}}
\put(96.00,28.00){\circle*{2.00}}
\put(95.00,34.00){\circle*{2.00}}
\put(97.00,44.00){\circle*{2.00}}
\put(99.00,50.00){\circle*{2.00}}
\put(117.00,15.00){\circle*{2.00}}
\put(122.00,20.00){\circle*{2.00}}
\put(120.00,26.00){\circle*{2.00}}
\put(120.00,10.00){\circle*{2.00}}
\put(117.00,4.00){\circle*{2.00}}
\put(128.00,20.00){\circle*{2.00}}
\put(127.00,14.00){\circle*{2.00}}
\put(133.00,12.00){\circle*{2.00}}
\put(132.00,8.00){\circle*{1.00}}
\put(140.00,16.00){\circle*{2.00}}
\put(145.00,15.00){\circle*{1.00}}
\put(146.00,25.00){\circle*{1.00}}
\put(141.00,23.00){\circle*{2.00}}
\put(135.00,25.00){\circle*{2.00}}
\put(135.00,30.00){\circle*{1.00}}
\put(110.00,37.00){\circle*{2.00}}
\put(107.00,33.00){\circle*{1.00}}
\put(108.00,51.00){\circle*{1.00}}
\put(122.00,54.00){\circle*{1.00}}
\put(117.00,38.00){\circle*{2.00}}
\put(119.00,34.00){\circle*{1.00}}
\put(122.00,45.00){\circle*{2.00}}
\put(127.00,46.00){\circle*{1.00}}
\put(133.00,49.00){\circle*{1.00}}
\put(143.00,54.00){\circle*{1.00}}
\put(143.00,48.00){\circle*{2.00}}
\put(137.00,46.00){\circle*{2.00}}
\put(138.00,40.00){\circle*{2.00}}
\put(135.00,39.00){\circle*{1.00}}
\put(145.00,36.00){\circle*{2.00}}
\put(146.00,31.00){\circle*{1.00}}
\put(149.00,44.00){\circle*{2.00}}
\put(150.00,38.00){\circle*{2.00}}
\put(157.00,51.00){\circle*{2.00}}
\put(156.00,45.00){\circle*{2.00}}
\put(161.00,39.00){\circle*{2.00}}
\put(159.00,35.00){\circle*{2.00}}
\put(161.00,29.00){\circle*{2.00}}
\put(152.00,20.00){\circle*{1.00}}
\put(157.00,22.00){\circle*{2.00}}
\put(160.00,11.00){\circle*{1.00}}
\put(169.00,15.00){\circle*{2.00}}
\put(171.00,29.00){\circle*{1.00}}
\put(181.00,31.00){\circle*{2.00}}
\put(185.00,26.00){\circle*{1.00}}
\put(190.00,35.00){\circle*{1.00}}
\put(179.00,42.00){\circle*{2.00}}
\put(182.00,47.00){\circle*{1.00}}
\put(168.00,47.00){\circle*{1.00}}
\put(169.00,38.00){\circle*{2.00}}
\put(184.00,7.00){\circle*{2.00}}
\put(192.00,13.00){\circle*{2.00}}
\put(191.00,19.00){\circle*{2.00}}
\put(196.00,22.00){\circle*{2.00}}
\put(196.00,27.00){\circle*{1.00}}
\put(203.00,17.00){\circle*{2.00}}
\put(204.00,10.00){\circle*{2.00}}
\put(199.00,9.00){\circle*{2.00}}
\put(199.00,4.00){\circle*{1.00}}
\put(208.00,18.00){\circle*{1.00}}
\put(210.00,8.00){\circle*{1.00}}
\put(100.00,40.00){\circle*{2.00}}
\emline{78.00}{38.00}{1}{77.00}{32.00}{2}
\emline{77.00}{32.00}{3}{82.00}{30.00}{4}
\emline{85.00}{43.00}{5}{78.00}{38.00}{6}
\emline{97.00}{44.00}{7}{100.00}{40.00}{8}
\emline{100.00}{40.00}{9}{95.00}{34.00}{10}
\emline{122.00}{45.00}{11}{117.00}{38.00}{12}
\emline{117.00}{38.00}{13}{110.00}{37.00}{14}
\emline{149.00}{44.00}{15}{150.00}{38.00}{16}
\emline{150.00}{38.00}{17}{145.00}{36.00}{18}
\emline{145.00}{36.00}{19}{138.00}{40.00}{20}
\emline{138.00}{40.00}{21}{137.00}{46.00}{22}
\emline{149.00}{44.00}{23}{156.00}{45.00}{24}
\emline{156.00}{45.00}{25}{161.00}{39.00}{26}
\emline{161.00}{39.00}{27}{159.00}{35.00}{28}
\emline{159.00}{35.00}{29}{150.00}{38.00}{30}
\emline{161.00}{39.00}{31}{169.00}{38.00}{32}
\put(172.00,43.00){\circle*{2.00}}
\put(184.00,36.00){\circle*{2.00}}
\emline{169.00}{38.00}{33}{172.00}{43.00}{34}
\emline{172.00}{43.00}{35}{179.00}{42.00}{36}
\emline{179.00}{42.00}{37}{184.00}{36.00}{38}
\emline{184.00}{36.00}{39}{181.00}{31.00}{40}
\emline{181.00}{31.00}{41}{174.00}{32.00}{42}
\emline{174.00}{32.00}{43}{169.00}{38.00}{44}
\put(174.00,32.00){\circle*{2.00}}
\emline{204.00}{10.00}{45}{199.00}{9.00}{46}
\emline{199.00}{9.00}{47}{192.00}{13.00}{48}
\emline{192.00}{13.00}{49}{191.00}{19.00}{50}
\emline{191.00}{19.00}{51}{196.00}{22.00}{52}
\emline{184.00}{7.00}{53}{184.00}{12.00}{54}
\emline{184.00}{12.00}{55}{192.00}{13.00}{56}
\emline{169.00}{15.00}{57}{172.00}{20.00}{58}
\emline{159.00}{35.00}{59}{161.00}{29.00}{60}
\emline{157.00}{51.00}{61}{156.00}{45.00}{62}
\emline{99.00}{50.00}{63}{97.00}{44.00}{64}
\emline{95.00}{34.00}{65}{96.00}{28.00}{66}
\emline{117.00}{15.00}{67}{122.00}{20.00}{68}
\emline{122.00}{20.00}{69}{128.00}{20.00}{70}
\emline{128.00}{20.00}{71}{127.00}{14.00}{72}
\emline{127.00}{14.00}{73}{120.00}{10.00}{74}
\emline{120.00}{10.00}{75}{117.00}{15.00}{76}
\emline{128.00}{20.00}{77}{135.00}{25.00}{78}
\emline{135.00}{25.00}{79}{141.00}{23.00}{80}
\emline{141.00}{23.00}{81}{140.00}{16.00}{82}
\emline{140.00}{16.00}{83}{133.00}{12.00}{84}
\emline{133.00}{12.00}{85}{127.00}{14.00}{86}
\emline{122.00}{20.00}{87}{120.00}{26.00}{88}
\emline{120.00}{10.00}{89}{117.00}{4.00}{90}
\emline{74.00}{39.00}{91}{78.00}{38.00}{92}
\emline{77.00}{32.00}{93}{73.00}{29.00}{94}
\emline{82.00}{30.00}{95}{82.00}{25.00}{96}
\emline{85.00}{43.00}{97}{85.00}{48.00}{98}
\emline{122.00}{45.00}{99}{127.00}{46.00}{100}
\emline{119.00}{34.00}{101}{117.00}{38.00}{102}
\emline{110.00}{37.00}{103}{107.00}{33.00}{104}
\emline{107.00}{33.00}{105}{107.00}{33.00}{106}
\emline{132.00}{8.00}{107}{133.00}{12.00}{108}
\emline{135.00}{25.00}{109}{135.00}{30.00}{110}
\emline{141.00}{23.00}{111}{146.00}{25.00}{112}
\emline{145.00}{15.00}{113}{140.00}{16.00}{114}
\emline{133.00}{49.00}{115}{137.00}{46.00}{116}
\emline{135.00}{39.00}{117}{138.00}{40.00}{118}
\emline{145.00}{36.00}{119}{146.00}{31.00}{120}
\emline{168.00}{47.00}{121}{172.00}{43.00}{122}
\emline{179.00}{42.00}{123}{182.00}{47.00}{124}
\emline{184.00}{36.00}{125}{190.00}{35.00}{126}
\emline{185.00}{26.00}{127}{181.00}{31.00}{128}
\emline{174.00}{32.00}{129}{171.00}{29.00}{130}
\emline{196.00}{27.00}{131}{196.00}{22.00}{132}
\emline{210.00}{8.00}{133}{204.00}{10.00}{134}
\emline{199.00}{4.00}{135}{199.00}{9.00}{136}
\put(184.00,12.00){\circle*{2.00}}
\emline{85.00}{43.00}{137}{90.00}{41.00}{138}
\emline{90.00}{41.00}{139}{89.00}{35.00}{140}
\emline{89.00}{35.00}{141}{82.00}{30.00}{142}
\emline{90.00}{41.00}{143}{97.00}{44.00}{144}
\put(90.00,41.00){\circle*{2.00}}
\put(89.00,35.00){\circle*{2.00}}
\emline{89.00}{35.00}{145}{95.00}{34.00}{146}
\put(107.00,41.00){\circle*{2.00}}
\emline{107.00}{41.00}{147}{100.00}{40.00}{148}
\put(112.00,48.00){\circle*{2.00}}
\emline{112.00}{48.00}{149}{107.00}{41.00}{150}
\put(119.00,49.00){\circle*{2.00}}
\emline{119.00}{49.00}{151}{112.00}{48.00}{152}
\emline{108.00}{51.00}{153}{112.00}{48.00}{154}
\emline{119.00}{49.00}{155}{122.00}{54.00}{156}
\emline{119.00}{49.00}{157}{122.00}{45.00}{158}
\emline{110.00}{37.00}{159}{107.00}{41.00}{160}
\put(110.00,14.00){\circle*{2.00}}
\put(98.00,6.00){\circle*{2.00}}
\put(95.00,2.00){\circle*{1.00}}
\put(96.00,20.00){\circle*{1.00}}
\put(110.00,23.00){\circle*{1.00}}
\put(105.00,7.00){\circle*{2.00}}
\put(107.00,3.00){\circle*{1.00}}
\put(110.00,14.00){\circle*{2.00}}
\emline{110.00}{14.00}{161}{105.00}{7.00}{162}
\emline{105.00}{7.00}{163}{98.00}{6.00}{164}
\emline{107.00}{3.00}{165}{105.00}{7.00}{166}
\emline{98.00}{6.00}{167}{95.00}{2.00}{168}
\put(95.00,10.00){\circle*{2.00}}
\put(100.00,17.00){\circle*{2.00}}
\emline{100.00}{17.00}{169}{95.00}{10.00}{170}
\put(107.00,18.00){\circle*{2.00}}
\emline{107.00}{18.00}{171}{100.00}{17.00}{172}
\emline{96.00}{20.00}{173}{100.00}{17.00}{174}
\emline{107.00}{18.00}{175}{110.00}{23.00}{176}
\emline{107.00}{18.00}{177}{110.00}{14.00}{178}
\emline{98.00}{6.00}{179}{95.00}{10.00}{180}
\emline{110.00}{14.00}{181}{117.00}{15.00}{182}
\emline{95.00}{10.00}{183}{90.00}{9.00}{184}
\emline{197.00}{49.00}{185}{188.00}{22.00}{186}
\emline{134.00}{59.00}{187}{125.00}{32.00}{188}
\emline{134.00}{59.00}{189}{197.00}{49.00}{190}
\emline{188.00}{22.00}{191}{125.00}{32.00}{192}
\emline{219.00}{28.00}{193}{210.00}{1.00}{194}
\emline{210.00}{1.00}{195}{147.00}{11.00}{196}
\emline{147.00}{11.00}{197}{152.00}{28.00}{198}
\emline{219.00}{28.00}{199}{191.00}{32.00}{200}
\put(172.00,20.00){\circle*{2.00}}
\emline{191.00}{19.00}{201}{187.00}{22.00}{202}
\emline{157.00}{22.00}{203}{159.00}{26.00}{204}
\emline{160.00}{27.00}{205}{159.00}{26.00}{206}
\emline{172.00}{20.00}{207}{169.00}{25.00}{208}
\put(90.00,9.00){\circle*{1.00}}
\put(172.00,10.00){\circle*{1.00}}
\emline{172.00}{10.00}{209}{169.00}{15.00}{210}
\emline{157.00}{22.00}{211}{152.00}{20.00}{212}
\emline{137.00}{46.00}{213}{143.00}{48.00}{214}
\emline{143.00}{48.00}{215}{149.00}{44.00}{216}
\emline{143.00}{48.00}{217}{143.00}{54.00}{218}
\emline{204.00}{10.00}{219}{203.00}{17.00}{220}
\emline{203.00}{17.00}{221}{196.00}{22.00}{222}
\emline{203.00}{17.00}{223}{208.00}{18.00}{224}
\emline{71.00}{47.00}{225}{132.00}{59.00}{226}
\emline{132.00}{59.00}{227}{131.00}{51.00}{228}
\emline{69.00}{19.00}{229}{130.00}{31.00}{230}
\emline{71.00}{47.00}{231}{67.00}{18.00}{232}
\emline{69.00}{19.00}{233}{67.00}{19.00}{234}
\emline{67.00}{18.00}{235}{70.00}{19.00}{236}
\emline{87.00}{-6.00}{237}{148.00}{6.00}{238}
\emline{89.00}{22.00}{239}{85.00}{-7.00}{240}
\emline{87.00}{-6.00}{241}{85.00}{-6.00}{242}
\emline{85.00}{-7.00}{243}{88.00}{-6.00}{244}
\emline{89.00}{22.00}{245}{134.00}{31.00}{246}
\emline{148.00}{6.00}{247}{149.00}{11.00}{248}
\emline{151.00}{25.00}{249}{151.00}{28.00}{250}
\emline{102.00}{14.00}{251}{136.00}{21.00}{252}
\emline{136.00}{21.00}{253}{136.00}{17.00}{254}
\emline{136.00}{17.00}{255}{102.00}{10.00}{256}
\emline{102.00}{10.00}{257}{102.00}{14.00}{258}
\emline{81.00}{38.00}{259}{115.00}{45.00}{260}
\emline{115.00}{45.00}{261}{115.00}{41.00}{262}
\emline{115.00}{41.00}{263}{81.00}{34.00}{264}
\emline{81.00}{34.00}{265}{81.00}{38.00}{266}
\emline{102.00}{14.00}{267}{102.00}{15.00}{268}
\emline{102.00}{15.00}{269}{98.00}{11.00}{270}
\emline{98.00}{11.00}{271}{102.00}{9.00}{272}
\emline{102.00}{9.00}{273}{102.00}{10.00}{274}
\emline{115.00}{46.00}{275}{119.00}{44.00}{276}
\emline{119.00}{44.00}{277}{115.00}{40.00}{278}
\emline{115.00}{45.00}{279}{115.00}{46.00}{280}
\emline{115.00}{41.00}{281}{115.00}{40.00}{282}
\emline{169.00}{15.00}{283}{161.00}{16.00}{284}
\emline{161.00}{16.00}{285}{157.00}{22.00}{286}
\emline{161.00}{16.00}{287}{160.00}{11.00}{288}
\put(161.00,16.00){\circle*{2.00}}
\put(179.00,19.00){\circle*{2.00}}
\emline{184.00}{12.00}{289}{179.00}{19.00}{290}
\emline{179.00}{19.00}{291}{172.00}{20.00}{292}
\emline{179.00}{19.00}{293}{182.00}{23.00}{294}
\emline{165.00}{23.00}{295}{198.00}{18.00}{296}
\emline{198.00}{18.00}{297}{198.00}{14.00}{298}
\emline{198.00}{14.00}{299}{164.00}{19.00}{300}
\emline{165.00}{24.00}{301}{165.00}{18.00}{302}
\emline{165.00}{18.00}{303}{161.00}{21.00}{304}
\emline{161.00}{21.00}{305}{165.00}{24.00}{306}
\emline{143.00}{44.00}{307}{177.00}{38.00}{308}
\emline{177.00}{34.00}{309}{142.00}{40.00}{310}
\emline{142.00}{40.00}{311}{143.00}{44.00}{312}
\emline{177.00}{33.00}{313}{179.00}{38.00}{314}
\emline{179.00}{38.00}{315}{182.00}{35.00}{316}
\emline{182.00}{35.00}{317}{177.00}{33.00}{318}
\emline{177.00}{38.00}{319}{179.00}{38.00}{320}
\put(0.00,21.00){\circle{13.00}}
\emline{8.00}{25.00}{321}{8.00}{17.00}{322}
\emline{8.00}{17.00}{323}{0.00}{12.00}{324}
\emline{0.00}{12.00}{325}{-8.00}{17.00}{326}
\emline{-8.00}{17.00}{327}{-8.00}{25.00}{328}
\emline{-8.00}{25.00}{329}{0.00}{30.00}{330}
\emline{0.00}{30.00}{331}{8.00}{25.00}{332}
\emline{8.00}{25.00}{333}{17.00}{28.00}{334}
\emline{17.00}{28.00}{335}{24.00}{21.00}{336}
\emline{24.00}{21.00}{337}{17.00}{14.00}{338}
\emline{17.00}{14.00}{339}{8.00}{17.00}{340}
\put(17.00,21.00){\circle{6.00}}
\emline{17.00}{28.00}{341}{17.00}{38.00}{342}
\emline{17.00}{14.00}{343}{17.00}{4.00}{344}
\put(44.00,21.00){\circle{10.00}}
\emline{36.00}{25.00}{345}{40.00}{29.00}{346}
\emline{40.00}{29.00}{347}{48.00}{29.00}{348}
\emline{48.00}{29.00}{349}{54.00}{21.00}{350}
\emline{54.00}{21.00}{351}{48.00}{13.00}{352}
\emline{48.00}{13.00}{353}{40.00}{13.00}{354}
\emline{40.00}{13.00}{355}{36.00}{17.00}{356}
\put(30.00,28.00){\circle{6.00}}
\put(17.00,1.00){\makebox(0,0)[cc]{O}}
\put(17.00,21.00){\makebox(0,0)[cc]{$-$}}
\put(30.00,28.00){\makebox(0,0)[cc]{$+$}}
\put(35.00,21.00){\makebox(0,0)[cc]{N}}
\put(17.00,41.00){\makebox(0,0)[cc]{O}}
\emline{18.00}{38.00}{357}{18.00}{36.00}{358}
\emline{18.00}{35.00}{359}{18.00}{33.00}{360}
\emline{18.00}{32.00}{361}{18.00}{30.00}{362}
\emline{18.00}{29.00}{363}{18.00}{28.00}{364}
\emline{18.00}{28.00}{365}{19.00}{27.00}{366}
\emline{20.00}{26.00}{367}{22.00}{24.00}{368}
\emline{23.00}{23.00}{369}{25.00}{21.00}{370}
\emline{25.00}{21.00}{371}{23.00}{19.00}{372}
\emline{22.00}{18.00}{373}{20.00}{16.00}{374}
\emline{19.00}{15.00}{375}{18.00}{14.00}{376}
\emline{18.00}{14.00}{377}{18.00}{13.00}{378}
\emline{18.00}{12.00}{379}{18.00}{10.00}{380}
\emline{18.00}{9.00}{381}{18.00}{7.00}{382}
\emline{18.00}{6.00}{383}{18.00}{4.00}{384}
\emline{32.00}{21.00}{385}{26.00}{21.00}{386}
\end{picture} \\

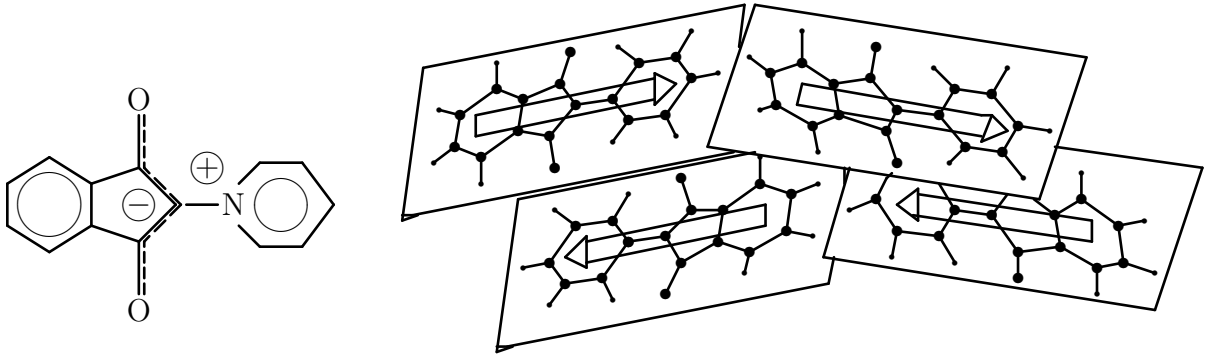
{\it Figure 5.  Structure  of  IPB  molecule (left).  Deposition of IPB
molecules in fundamental crystal unit (right).  The  electric  dipole
moments  of  the  IPB  molecules are shown that demonstrates it vortex
(toroid) distribution.}
\vskip 1mm
\noindent
Let us see one more example. In a cubic perovskite the crystallographic 
plane $(1,1,1)$ contains triangular sublattice of oxygen atoms [21,- 23]
(see Figure 6). 
Magnetic moments of these atoms can be in two states of orientating 
symmetry that are separated by temperatural phase transition. \\
\noindent
\begin{minipage}{8cm}
\unitlength=1mm
\special{em:linewidth 0.4pt}
\linethickness{0.4pt}
\begin{picture}(34.00,55.00)
\put(17.00,39.00){\circle*{2.00}}
\put(37.00,39.00){\circle*{2.00}}
\put(37.00,19.00){\circle*{2.00}}
\put(17.00,19.00){\circle*{2.00}}
\put(22.00,24.00){\circle*{2.00}}
\put(42.00,24.00){\circle*{2.00}}
\put(42.00,44.00){\circle*{2.00}}
\put(17.00,39.00){\line(1,0){20.00}}
\put(37.00,39.00){\line(1,1){5.00}}
\put(22.00,44.00){\circle*{2.00}}
\put(17.00,39.00){\line(1,1){5.00}}
\put(22.00,44.00){\line(1,0){20.00}}
\put(42.00,24.00){\line(-1,-1){5.00}}
\put(37.00,19.00){\line(-1,0){20.00}}
\put(17.00,19.00){\line(1,1){5.00}}
\put(22.00,24.00){\line(1,0){20.00}}
\put(42.00,24.00){\line(0,1){20.00}}
\put(37.00,39.00){\line(0,-1){20.00}}
\put(17.00,19.00){\line(0,1){20.00}}
\put(22.00,24.00){\line(0,1){20.00}}
\put(22.00,44.00){\line(-1,-5){5.00}}
\put(17.00,19.00){\line(5,1){25.00}}
\put(42.00,24.00){\line(-1,1){20.00}}
\put(20.00,32.00){\circle{4.00}}
\put(29.00,22.00){\circle{4.00}}
\put(33.00,33.00){\circle{4.00}}
\put(30.00,41.00){\circle{4.00}}
\put(27.00,28.00){\circle{4.00}}
\put(40.00,32.00){\circle{4.00}}
\put(30.00,30.00){\circle*{2.00}}
\put(54.00,32.00){\circle{4.00}}
\put(54.00,41.00){\circle*{2.00}}
\put(30.00,30.00){\circle*{4.00}}
\put(54.00,22.00){\circle*{4.00}}
\put(59.00,38.00){\makebox(9.00,5.00)[cc]{$Ba^{2+}$}}
\put(59.00,28.00){\makebox(9.00,5.00)[cc]{$O^{2-}$}}
\put(59.00,19.00){\makebox(9.00,5.00)[cc]{$Ti^{4+}$}}
\end{picture}
\end{minipage} 
\begin{minipage}{8cm}
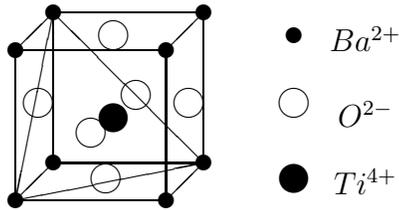
{\it Figure 6. Structure of type perovskite-type compound 
Barium Titanate $BaTiO_3$. 
In reality atoms fill up the major part of space. Here the position of 
their center is shown only.} \end{minipage} \\ \noindent \begin{minipage}{8cm} 
\unitlength=1.00mm \special{em:linewidth 0.4pt} \linethickness{0.4pt} 
\begin{picture}(140.00,72.00) \put(20.00,70.00){\circle{4.00}} 
\put(40.00,70.00){\circle{4.00}}
\put(60.00,70.00){\circle{4.00}}
\put(65.00,57.00){\circle{4.00}}
\put(55.00,57.00){\circle{4.00}}
\put(45.00,57.00){\circle{4.00}}
\put(35.00,57.00){\circle{4.00}}
\put(25.00,57.00){\circle{4.00}}
\put(15.00,57.00){\circle{4.00}}
\put(30.00,45.00){\circle{4.00}}
\put(50.00,45.00){\circle{4.00}}
\put(65.00,32.00){\circle{4.00}}
\put(55.00,32.00){\circle{4.00}}
\put(45.00,32.00){\circle{4.00}}
\put(35.00,32.00){\circle{4.00}}
\put(25.00,32.00){\circle{4.00}}
\put(15.00,32.00){\circle{4.00}}
\put(20.00,20.00){\circle{4.00}}
\put(40.00,20.00){\circle{4.00}}
\put(60.00,20.00){\circle{4.00}}
\put(10.00,70.00){\circle*{2.00}}
\put(30.00,70.00){\circle*{2.00}}
\put(50.00,70.00){\circle*{2.00}}
\put(70.00,70.00){\circle*{2.00}}
\put(60.00,45.00){\circle*{2.00}}
\put(40.00,45.00){\circle*{2.00}}
\put(20.00,45.00){\circle*{2.00}}
\put(10.00,20.00){\circle*{2.00}}
\put(30.00,20.00){\circle*{2.00}}
\put(50.00,20.00){\circle*{2.00}}
\put(70.00,20.00){\circle*{2.00}}
\put(15.00,70.00){\vector(1,0){10.00}}
\put(35.00,70.00){\vector(1,0){10.00}}
\put(55.00,70.00){\vector(1,0){10.00}}
\put(17.00,52.00){\vector(-1,3){3.67}}
\put(27.00,63.00){\vector(-1,-3){4.00}}
\put(25.00,45.00){\vector(1,0){10.00}}
\put(37.00,52.00){\vector(-1,3){3.33}}
\put(47.00,62.00){\vector(-1,-2){5.67}}
\put(46.00,45.00){\vector(1,0){9.00}}
\put(57.00,52.00){\vector(-1,3){3.67}}
\put(66.00,61.00){\vector(-1,-3){3.33}}
\put(57.00,37.00){\vector(-1,-3){4.00}}
\put(55.00,20.00){\vector(1,0){10.00}}
\put(67.00,27.00){\vector(-1,3){3.33}}
\put(47.00,27.00){\vector(-1,3){3.67}}
\put(37.00,38.00){\vector(-1,-3){4.00}}
\put(35.00,20.00){\vector(1,0){10.00}}
\put(27.00,27.00){\vector(-1,3){3.67}}
\put(17.00,38.00){\vector(-1,-3){4.33}}
\put(15.00,20.00){\vector(1,0){10.00}}
\end{picture}
\end{minipage}
\begin{minipage}{8cm}
\unitlength=1.00mm
\special{em:linewidth 0.4pt}
\linethickness{0.4pt}
\begin{picture}(73.00,72.00)
\put(10.00,20.00){\line(2,5){20.00}}
\put(50.00,70.00){\line(-2,-5){20.00}}
\put(50.00,20.00){\line(2,5){20.00}}
\put(50.00,70.00){\line(2,-5){20.00}}
\put(50.00,20.00){\line(-2,5){20.00}}
\put(10.00,70.00){\line(2,-5){20.00}}
\put(6.00,20.00){\line(1,0){67.00}}
\put(73.00,45.00){\line(-1,0){67.00}}
\put(6.00,70.00){\line(1,0){67.00}}
\put(20.00,70.00){\circle{4.00}}
\put(40.00,70.00){\circle{4.00}}
\put(60.00,70.00){\circle{4.00}}
\put(65.00,57.00){\circle{4.00}}
\put(55.00,57.00){\circle{4.00}}
\put(45.00,57.00){\circle{4.00}}
\put(35.00,57.00){\circle{4.00}}
\put(25.00,57.00){\circle{4.00}}
\put(15.00,57.00){\circle{4.00}}
\put(30.00,45.00){\circle{4.00}}
\put(50.00,45.00){\circle{4.00}}
\put(65.00,32.00){\circle{4.00}}
\put(55.00,32.00){\circle{4.00}}
\put(45.00,32.00){\circle{4.00}}
\put(35.00,32.00){\circle{4.00}}
\put(25.00,32.00){\circle{4.00}}
\put(15.00,32.00){\circle{4.00}}
\put(20.00,20.00){\circle{4.00}}
\put(40.00,20.00){\circle{4.00}}
\put(60.00,20.00){\circle{4.00}}
\put(10.00,70.00){\circle*{2.00}}
\put(30.00,70.00){\circle*{2.00}}
\put(50.00,70.00){\circle*{2.00}}
\put(70.00,70.00){\circle*{2.00}}
\put(60.00,45.00){\circle*{2.00}}
\put(40.00,45.00){\circle*{2.00}}
\put(20.00,45.00){\circle*{2.00}}
\put(10.00,20.00){\circle*{2.00}}
\put(30.00,20.00){\circle*{2.00}}
\put(50.00,20.00){\circle*{2.00}}
\put(70.00,20.00){\circle*{2.00}}
\put(15.00,70.00){\vector(1,0){10.00}}
\put(35.00,70.00){\vector(1,0){10.00}}
\put(55.00,70.00){\vector(1,0){10.00}}
\put(17.00,52.00){\vector(-1,3){3.67}}
\put(27.00,63.00){\vector(-1,-3){4.00}}
\put(25.00,45.00){\vector(1,0){10.00}}
\put(37.00,52.00){\vector(-1,3){3.33}}
\put(47.00,62.00){\vector(-1,-2){5.67}}
\put(46.00,45.00){\vector(1,0){9.00}}
\put(57.00,52.00){\vector(-1,3){3.67}}
\put(66.00,61.00){\vector(-1,-3){3.33}}
\put(57.00,37.00){\vector(-1,-3){4.00}}
\put(55.00,20.00){\vector(1,0){10.00}}
\put(67.00,27.00){\vector(-1,3){3.33}}
\put(47.00,27.00){\vector(-1,3){3.67}}
\put(37.00,38.00){\vector(-1,-3){4.00}}
\put(35.00,20.00){\vector(1,0){10.00}}
\put(27.00,27.00){\vector(-1,3){3.67}}
\put(17.00,38.00){\vector(-1,-3){4.33}}
\put(15.00,20.00){\vector(1,0){10.00}}
\end{picture}
\end{minipage}

\noindent
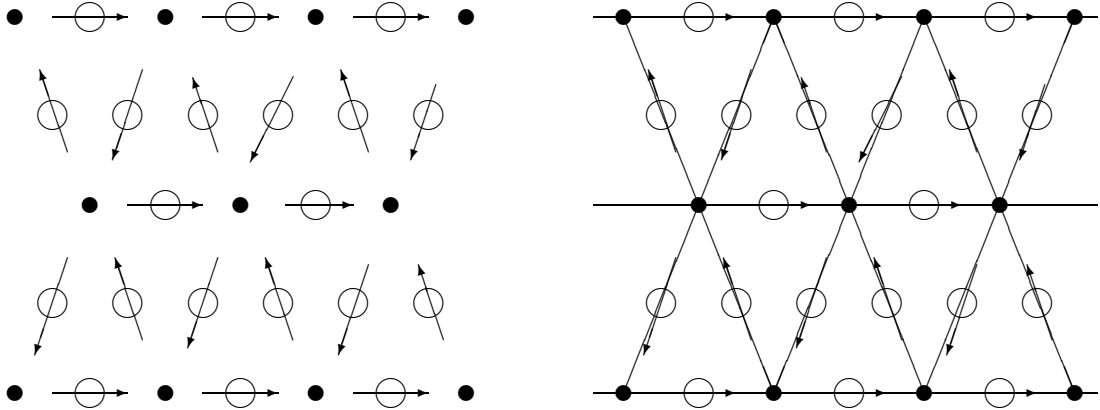
{\it Figure 7a.
Magnetic moments of oxygen atoms are orientated in such a 
way that they generate in each triangle toroid moment directed "up" or 
"down" by turns and thus build up anti-toroic}
\vskip 2mm
\noindent
\begin{minipage}{8cm}
\unitlength=1.00mm
\special{em:linewidth 0.4pt}
\linethickness{0.4pt}
\begin{picture}(71.00,74.00)
\put(20.00,70.00){\circle{4.00}}
\put(40.00,70.00){\circle{4.00}}
\put(60.00,70.00){\circle{4.00}}
\put(65.00,57.00){\circle{4.00}}
\put(55.00,57.00){\circle{4.00}}
\put(45.00,57.00){\circle{4.00}}
\put(35.00,57.00){\circle{4.00}}
\put(25.00,57.00){\circle{4.00}}
\put(15.00,57.00){\circle{4.00}}
\put(30.00,45.00){\circle{4.00}}
\put(50.00,45.00){\circle{4.00}}
\put(65.00,32.00){\circle{4.00}}
\put(55.00,32.00){\circle{4.00}}
\put(45.00,32.00){\circle{4.00}}
\put(35.00,32.00){\circle{4.00}}
\put(25.00,32.00){\circle{4.00}}
\put(15.00,32.00){\circle{4.00}}
\put(20.00,20.00){\circle{4.00}}
\put(40.00,20.00){\circle{4.00}}
\put(60.00,20.00){\circle{4.00}}
\put(10.00,70.00){\circle*{2.00}}
\put(30.00,70.00){\circle*{2.00}}
\put(50.00,70.00){\circle*{2.00}}
\put(70.00,70.00){\circle*{2.00}}
\put(60.00,45.00){\circle*{2.00}}
\put(40.00,45.00){\circle*{2.00}}
\put(20.00,45.00){\circle*{2.00}}
\put(10.00,20.00){\circle*{2.00}}
\put(30.00,20.00){\circle*{2.00}}
\put(50.00,20.00){\circle*{2.00}}
\put(70.00,20.00){\circle*{2.00}}
\put(20.00,74.00){\vector(0,-1){10.00}}
\put(40.00,74.00){\vector(0,-1){10.00}}
\put(60.00,74.00){\vector(0,-1){10.00}}
\put(69.00,54.00){\vector(-3,2){9.00}}
\put(51.00,54.00){\vector(4,3){7.00}}
\put(48.00,55.00){\vector(-2,1){7.00}}
\put(31.00,54.00){\vector(3,2){8.00}}
\put(29.00,55.00){\vector(-2,1){8.00}}
\put(12.00,55.00){\vector(2,1){7.00}}
\put(30.00,49.00){\vector(0,-1){10.00}}
\put(50.00,49.00){\vector(0,-1){10.00}}
\put(59.00,29.00){\vector(-3,2){8.00}}
\put(41.00,29.00){\vector(3,2){8.00}}
\put(38.00,30.00){\vector(-2,1){7.00}}
\put(21.00,29.00){\vector(3,2){8.00}}
\put(18.00,29.00){\vector(-4,3){7.00}}
\put(20.00,25.00){\vector(0,-1){11.00}}
\put(40.00,25.00){\vector(0,-1){11.00}}
\put(60.00,25.00){\vector(0,-1){11.00}}
\put(62.00,29.00){\vector(4,3){7.00}}
\end{picture}
\end{minipage}
\begin{minipage}{8cm}
\unitlength=1.00mm
\special{em:linewidth 0.4pt}
\linethickness{0.4pt}
\begin{picture}(73.00,74.00)
\put(10.00,20.00){\line(2,5){20.00}}
\put(50.00,70.00){\line(-2,-5){20.00}}
\put(50.00,20.00){\line(2,5){20.00}}
\put(50.00,70.00){\line(2,-5){20.00}}
\put(50.00,20.00){\line(-2,5){20.00}}
\put(10.00,70.00){\line(2,-5){20.00}}
\put(6.00,20.00){\line(1,0){67.00}}
\put(73.00,45.00){\line(-1,0){67.00}}
\put(6.00,70.00){\line(1,0){67.00}}
\put(20.00,70.00){\circle{4.00}}
\put(40.00,70.00){\circle{4.00}}
\put(60.00,70.00){\circle{4.00}}
\put(65.00,57.00){\circle{4.00}}
\put(55.00,57.00){\circle{4.00}}
\put(45.00,57.00){\circle{4.00}}
\put(35.00,57.00){\circle{4.00}}
\put(25.00,57.00){\circle{4.00}}
\put(15.00,57.00){\circle{4.00}}
\put(30.00,45.00){\circle{4.00}}
\put(50.00,45.00){\circle{4.00}}
\put(65.00,32.00){\circle{4.00}}
\put(55.00,32.00){\circle{4.00}}
\put(45.00,32.00){\circle{4.00}}
\put(35.00,32.00){\circle{4.00}}
\put(25.00,32.00){\circle{4.00}}
\put(15.00,32.00){\circle{4.00}}
\put(20.00,20.00){\circle{4.00}}
\put(40.00,20.00){\circle{4.00}}
\put(60.00,20.00){\circle{4.00}}
\put(10.00,70.00){\circle*{2.00}}
\put(30.00,70.00){\circle*{2.00}}
\put(50.00,70.00){\circle*{2.00}}
\put(70.00,70.00){\circle*{2.00}}
\put(60.00,45.00){\circle*{2.00}}
\put(40.00,45.00){\circle*{2.00}}
\put(20.00,45.00){\circle*{2.00}}
\put(10.00,20.00){\circle*{2.00}}
\put(30.00,20.00){\circle*{2.00}}
\put(50.00,20.00){\circle*{2.00}}
\put(70.00,20.00){\circle*{2.00}}
\put(20.00,74.00){\vector(0,-1){10.00}}
\put(40.00,74.00){\vector(0,-1){10.00}}
\put(60.00,74.00){\vector(0,-1){10.00}}
\put(69.00,54.00){\vector(-3,2){9.00}}
\put(51.00,54.00){\vector(4,3){7.00}}
\put(48.00,55.00){\vector(-2,1){7.00}}
\put(31.00,54.00){\vector(3,2){8.00}}
\put(29.00,55.00){\vector(-2,1){8.00}}
\put(12.00,55.00){\vector(2,1){7.00}}
\put(30.00,49.00){\vector(0,-1){10.00}}
\put(50.00,49.00){\vector(0,-1){10.00}}
\put(59.00,29.00){\vector(-3,2){8.00}}
\put(41.00,29.00){\vector(3,2){8.00}}
\put(38.00,30.00){\vector(-2,1){7.00}}
\put(21.00,29.00){\vector(3,2){8.00}}
\put(18.00,29.00){\vector(-4,3){7.00}}
\put(20.00,25.00){\vector(0,-1){11.00}}
\put(40.00,25.00){\vector(0,-1){11.00}}
\put(60.00,25.00){\vector(0,-1){11.00}}
\put(62.00,29.00){\vector(4,3){7.00}}
\end{picture}
\end{minipage}
\noindent
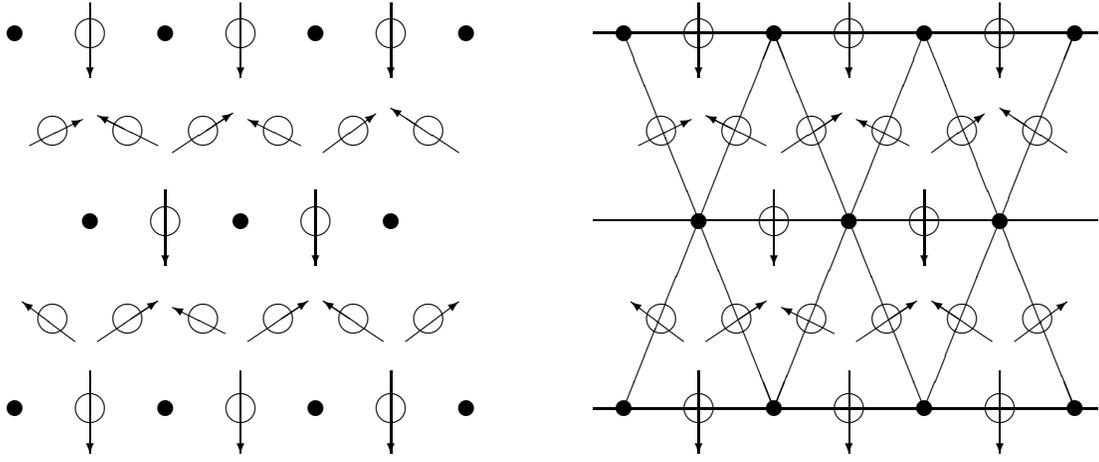
{\it Figure 7b. 
 Triangular anti-regulation of meansquare radii of 
magnetic moments of each domain in $(1,1,1)$ plane.}

\vskip 5mm 
\section{Conclusion} \noindent Now let us remember that 
there are two electric fields those differ by properties [9].  That, which 
appears in transverse part, is radiation field, unlike the longitudinal 
part that can be connected with the evolution of scalar one in Lorentz 
gauge \begin{equation} \mbox{div}\bA\,=\,-\frac{1}{c}\,\dot \varphi.  
\end{equation} In connection with this we should notice the terminological 
ambiguity.  In the framework of electrodynamics of continuous media, where 
may be realized the situation when $\mbox{div}\bj\,=\,0$, the moments 
$\dot Q_{lm}$ are the functions, independent to Coulomb moments $Q_{lm}$, 
because there is no free charges in the system considered. So $Q_{lm}$ 
cannot be restored as a result of the measurement of Coulomb moments by 
means of permanent electric field.  We will illustrate it doing inverse 
transformation, i.e.  adding in the current part of multipole expansion 
the transverse and longitudinal contributions, for simplicity ${\dot 
Q}_{1m}$:  \begin{eqnarray} l^{-1} \bigl[\dot 
Q_{lm}\triangle^{-1}Y_{lm}(r\dot {\bD})\,-\,\dot Q_{lm}Y_{lm} 
({\nabla})\triangle^{n-1}\mbox{div}\,\bA\bigr]_{l=1,n=0}\Longrightarrow 
\nonumber\\ 
\Longrightarrow \dot {\bQ} \triangle^{-1}\bigl(\frac{1}{c}\dot {\bD} - 
\nabla \mbox{div}\bA\bigr)\,=\,\dot {\bQ} 
\triangle^{-1}\bigl(\mbox{curl\,curl} \bA \,-\, \nabla 
\mbox{div}\bA\bigr)\,=\, \nonumber   \\
=\,-\dot {\bQ} \triangle^{-1}\triangle \bA \,= \,-\dot {\bQ} \,\bA 
\,=\,\bQ \dot {\bA} -\frac{d }{dt}\bQ\bA\,\Longrightarrow\,- \bQ 
 \bE \end{eqnarray} 
Here we take into account that $\frac{1}{c} \dot 
{\bE}\,=\,\mbox{curl}\, \bB \,=\, \mbox{curl\,curl}\,\bA$.  Naturally, 
these expressions are nonzero and can be justified only if 
the system is described but by a charge density and 
$\dot{\bQ}\,=\,\int\,\bj\,d\br$ (see e.g. [9]). Note that in quantum 
mechanics of atoms and moluclues there are always isolated charges and the 
moments $\dot{Q}_{lm}$ and $Q_{lm}$ are connected by the evolution 
equation (3.19) i.e. $\dot{Q}_{lm} \,=\,i[H_c,Q_{lm}]$.

The Coulomb gauge is generally not applied if a system in consideration 
contains some free charges. Therefore we give the multipole 
expansion of $\rho(\br,\,t)$ for the completeness of consideration [11, 3] 
\begin{equation} \rho(\br,t)\,=\,\sum_{l,m,n=0} \frac{(2l+1)!!}{2^n n!  
(2l+2n+1)!!}\, \sqrt{\frac{4\pi}{2l+1}} 
Q_{lm}^{(2n)}(t)\triangle^n\delta_{lm}(\br), \end{equation} 
\begin{equation} 
Q_{lm}(t)\,=\,\sqrt{\frac{4\pi}{2l+1}}\int\,r^{l+2n}Y_{lm}^{*}(\hat r)
\rho(\br, t)\,d\br. 
\end{equation}
Naturally, the Coulomb dipole moments, inherent to (4.24), 
interact with electric field as usual $H\,=\,-\bQ \bE$.
Had we used the Dirac's analysis of constrained Hamiltonian system 
developed within the scope of our problem in [24,\,25] we would have 
been in need of the fixation of Coulomb gauge [26] (see also [9]).  

\vskip 5mm 
\noindent
{\bf Acknowledgements}

\noindent
B. Saha Would like to thank Prof. S. Randjbar-Daemi and ICTP High Energy 
Section for hospitality. 
\vskip 5mm
\section{Appendix}
\noindent
According to definition the current density, corresponding to one 
dipole moment in the proper system of coordinates (see Fig. 1) is equal
(see e.g. [13])
$$\bJ^{\alpha}(\bq^{\alpha},\,\bR)\,=\,\dot{\bQ}^{\alpha}\,
\delta(\bq^{\alpha}\,-\,\bR). \eqno(A.1) $$
where value of dipole is defined as 
$\bQ^{\alpha}\,=\,-e\,(\bq^{\alpha}\,-\,\bR).$ 
We may introduce the total point-like current 
$\sum_{\alpha}\dot{\bq}^{\alpha}$ considered to be localized in the fixed 
point $R$ in the coordinate system $\br$ and attach it to the fixed point 
$\bR$ (Fig.  1):  
$$\bJ(\br)\,:=\,\sum_{\alpha}\dot{\bq}^{\alpha}\,\delta(\br\,-\,\bR), 
\eqno(A.2)$$
where the proper dipoles $\bQ^{\alpha}\,=\,-e\,(\bq^{\alpha}\,-\,\bR)$
as before, but $\dot{\bQ}^{\alpha}\,=\,\dot{\bq}^{\alpha}.$ Certainly we 
may obtain the last definition from the exact definition of the total 
current density of all electrons in the system of coordinates $\br$ using 
dipole approximation:  
$$ 
\bJ(\br)\,=\,\sum_{\alpha}\dot{\bq}^{\alpha}\delta(\br-\bq^{\alpha}) 
\approx \sum_{\alpha}\dot{\bq}^{\alpha}\delta(\br-\bR),\eqno(A.3) $$ 
as far as 
$$\delta(\br - 
\bq^{\alpha})\,=\,\delta(\br-\bR+\bR-\bq^{\alpha})|_{\bR\to \bq^{\alpha}} 
\approx \delta(\br-\bR)+(\bR-\bq^{\alpha})\nabla\delta(\br-\bR)+\cdots 
$$ 
But in the system $\br$, the polarization vector field of the molecular 
system can be represented as [14] 
$$\bP(\br,\,\bq^{\alpha})\Longrightarrow 
\sum_{\alpha}\bp(\bq^{\alpha})\delta(\br-\bq^{\alpha})\,=\, 
\sum_{\alpha}\bp(\bq^{\alpha})\delta(\br-\bR+\bR-\bq^{\alpha})|_
{\bR \to \bq^{\alpha}}\approx\sum_{\alpha}
\bp(\bq^{\alpha})\delta(\br-\bR)+\cdots  \eqno(A.4)$$
where $\sum_{\alpha}\bp^{\alpha}$ is the sum of dipoles of all 
electron. From here the polarization current is found as 
$$ \bJ(\br)\,:=\,\frac{\partial \bP}{\partial t}\Longrightarrow 
\sum_{\alpha}\dot{\bp}(\bq^{\alpha}) \delta(\br-\bR). \eqno(A.5)$$ 
Comparing (A.5) with (A.2) we conclude that
$$\bP(\br,\,\bq^{\alpha})\,=\,-e\,(\bq^{\alpha}\,-\,\bR)\,
\delta(\br\,-\,\bR)\,+\cdots \eqno(A.6)$$
\noindent
Now to define the quadrupole moment we use the following expression
$$Q_{ij}\,=\,\frac{1}{2}\int\,\Bigl[\bP_i\br_j\,+\,\br_i\bP_j\,-\,
\frac{2}{3}\bigl(\br\cdot\bP\bigr)\delta_{ij}\Bigr]\,d\br. \eqno(A.7)$$
Putting the value $\bP$ from (A.6) we obtain
$$ Q_{ij}\,=\,-\frac{e}{2}\Bigl[\bq_{i}^{\alpha}\br_j + \br_i 
\bq_{j}^{\alpha} - 2 \br_i\br_j + \frac{2}{3}\br(\bq - \br)\delta_{ij}
\Bigr].
\eqno(A.8)$$

\end{document}